\documentclass[%
 reprint,
 superscriptaddress,
 amsmath,amssymb,
 aps,
 nature,
]{revtex4-1}

\usepackage{graphicx}
\usepackage{dcolumn}
\usepackage{bm}
\usepackage{xcolor}
\usepackage{algorithm}
\usepackage{algpseudocode}
\usepackage{url} 
\usepackage{makecell}
\usepackage{multirow}
\usepackage{booktabs}
\usepackage{graphicx}
\usepackage{subfig}
\usepackage{comment}
\usepackage{braket}
\usepackage[amssymb]{SIunits}
\usepackage{bibunits}
\usepackage{booktabs,tabularx}
\usepackage[normalem]{ulem}

\begin{document}

\title{Non-collinear Magnetic Atomic Cluster Expansion for Iron}

\author{Matteo Rinaldi}
\email{matteo.rinaldi@rub.de}
\affiliation{Interdisciplinary Centre for Advanced Materials Simulation, Ruhr-Universit{\"a}t Bochum, 44801 Bochum, Germany}
\author{Matous Mrovec}
\email{matous.mrovec@rub.de}
\affiliation{Interdisciplinary Centre for Advanced Materials Simulation, Ruhr-Universit{\"a}t Bochum, 44801 Bochum, Germany}
\author{Anton Bochkarev}
\email{anton.bochkarev@rub.de}
\affiliation{Interdisciplinary Centre for Advanced Materials Simulation, Ruhr-Universit{\"a}t Bochum, 44801 Bochum, Germany}
\author{Yury Lysogorskiy}
\email{yury.lysogorskiy@icams.rub.de}
\affiliation{Interdisciplinary Centre for Advanced Materials Simulation, Ruhr-Universit{\"a}t Bochum, 44801 Bochum, Germany}
\author{Ralf Drautz}
\email{ralf.drautz@rub.de}
\affiliation{Interdisciplinary Centre for Advanced Materials Simulation, Ruhr-Universit{\"a}t Bochum, 44801 Bochum, Germany}

\date{\today}

\begin{abstract}

The Atomic Cluster Expansion (ACE) provides a formally complete basis for the local atomic environment. 
ACE is not limited to representing energies as a function of atomic positions and chemical species, but can be generalized to vectorial or tensorial properties and to incorporate further degrees of freedom (DOF). This is crucial for magnetic materials with potential energy surfaces that depend on atomic positions and atomic magnetic moments simultaneously. In this work, we employ the ACE formalism to develop a non-collinear magnetic ACE parametrization for the prototypical magnetic element Fe. 
The model is trained on a broad range of collinear and non-collinear magnetic structures calculated using spin density functional theory. We demonstrate that the non-collinear magnetic ACE is able to reproduce not only ground state properties of various magnetic phases of Fe but also the magnetic and lattice excitations that are essential for a correct description of the finite temperature behavior and properties of crystal defects.

\end{abstract}

\maketitle

\section{Introduction}

Recent advancements of data-driven methods and machine-learned (ML) interatomic potentials have led to dramatically improved descriptions of the potential energy surface (PES) for many material systems. However, the incorporation of spin degrees of freedom (DOF), which are crucial to capture finite temperature phenomena in magnetic materials, has remained a challenging endeavour. In spin density functional theory (SDFT), magnetizaton emerges from the competition of magnetic exchange and band energy contributions~\cite{UvonBarth_1972,stoner1938collective}, where the energy required for reshuffling electrons in up and down spin channels depends on the local density of states (DOS). The bimodal DOS of iron in the body-centred crystal (bcc) structure affords large DOS values close to the Fermi level, leading to larger magnetic moments than in the face-centred cubic (fcc) structure with its more unimodal DOS that is lower at the Fermi level \cite{Pettiforbook,Drautz2006BOP}. This intricate interplay between magnetic and atomic structure implies that multi-atom multi-spin interactions are necessary for capturing different magnetic and atomic structures in a single model.        

Unlike approaches that were derived from electronic structure theory and that seamlessly incorporate the complexity of magnetic interactions~\cite{soulairol2010structure, drautz2011valence, mrovec2011magnetic}, classical interatomic potentials needed to be supplemented via suitable interaction terms that mimic the quantum exchange interactions. The simplest possibility was to employ a classical Heisenberg Hamiltonian~\cite{heisenberg}, where the atomic spin operators are substituted by spin vectors and the exchange interactions are parameterized using first-principles calculations~\cite{TRANCHIDA2018406}. Such strategies have been adopted also in most existing ML approaches for magnetic systems.

Nikolov et al.~\cite{nikolov2021data} augmented the spectral neighborhood analysis potential (SNAP) framework with a two-spin bi-linear Heisenberg model with atomic magnetic moment magnitudes being fixed and independent of the environment. A similar approach, where a neural network was trained to describe contributions to the Heisenberg Hamiltonian based on the local magnetic environment, was developed by Yu et al.~\cite{Yu_PhysRevB.105.174422_2022}. However, this approach did not include information about the underlying lattice and treated the magnetic moments as unit vectors. Eckhoff et al.~\cite{Eckhoff2021HighdimensionalNN} extended the formalism based on Behler-Parrinello symmetry functions~\cite{PhysRevLett.98.146401} in a framework that was limited to collinear configurations. Magnetic moments as additional DOF were incorporated  by Novikov et al.~\cite{novikov2022magnetic} in the moment tensor potential framework~\cite{shapeev2016moment}. Even though the description was confined to collinear moments only, the magnetic moment tensor potential was able to reproduce a number of thermodynamic properties of bulk bcc Fe. Very recently, Domina et al.~\cite{Domina_PhysRevB.105.214439_2022} extended the SNAP framework to deal with arbitrary vectorial fields and demonstrated its functionality by training to non-collinear spin configurations generated using a model Landau-Heisenberg Hamiltonian. Finally, aiming at large-scale spin-lattice dynamics simulations, Chapman et al.~\cite{chapman2022machine} added a neural network correction term to an embedded atom method potential augmented with a Heisenberg-Landau Hamiltonian. The model was successfully applied in finite temperature simulations of bulk Fe phases as well as complex defects. However, due to its simplicity, absolute errors were in some cases larger than a few tens of meV that are comparable to the fluctuations of exchange parameters with temperature.  Thus, none of the existing magnetic ML approaches has so far succeeded in achieving a transferable and quantitatively accurate description of magnetic interactions suitable for modelling magnetism in different crystal structures.


We present an explicit treatment of non-collinear magnetic DOF within the atomic cluster expansion (ACE)~\cite{Drautz_PhysRevB.99.014104, Drautz_PhysRevB.102.024104}. ACE provides a complete basis in the space of atomic environments \cite{Drautz_PhysRevB.99.014104,dusson2022atomic} and accurate, transferable and numerically efficient parameterizations of ACE have been developed for diverse bonding environments including bulk metallic systems as well as covalent molecules \cite{lysogorskiy2021performant,Kovacs21,qamar2022atomic}. Thanks to ACE universality, additional scalar, vectorial or tensorial DOF can be incorporated seamlessly into ACE models~\cite{Drautz_PhysRevB.102.024104}. Specifically for magnetic systems, ACE provides a body-ordered decomposition of combined atomic and magnetic PES in terms of a complete set of basis functions that depend on atomic and magnetic DOF. The inclusion of magnetic DOF requires an extension of the ACE equivariant basis such that any transformation of the relevant translation and rotation symmetry group acting on both atomic and magnetic spaces leaves the energy invariant. Magnetic ACE can therefore be considered as a generalization of most existing magnetic ML models as well as the classical spin-cluster expansion (SCE)~\cite{SingerFahnle,PhysRevLett.107.017204,PhysRevB.69.104404,PhysRevB.72.212405,PhysRevB.81.184202}.

In this work, we develop a non-collinear magnetic ACE parameterization for the prototypical magnetic element Fe. The model is trained on a large dataset of both collinear and non-collinear DFT calculations and validated for a broad range of structural, thermodynamic and defect properties. The resulting interatomic potential is able to describe accurately complex potential energy landscapes of different magnetic and atomic phases of Fe as a function of both atomic positions and local magnetic moment vectors.

\section{Results and Discussion}

\subsection{Reference DFT data \label{sec:DFT}}

A comprehensive sampling of variations in both atomic positions and magnetic moments is crucial for the construction of any atomistic magnetic ML model. Sampling of the atomic DOF can be carried out following well established protocols employed in ML fitting of PES, commonly by choosing a set of structures and varying their geometries and atomic positions. In contrast, sampling of the magnetic DOF presents a significant difficulty, both from the computational and methodological point of view. Firstly, the number of required calculations increases drastically due to the additional spin degrees of freedom and, secondly, the local atomic magnetic moments need to be constrained to desired magnitudes and orientations~\cite{PhysRevB.91.054420}. While it is in principle possible to fix both the direction and the magnitude of each atomic magnetic moment to a target vector~\cite{PhysRevB.91.054420}, these calculations are computationally demanding. Furthermore, as atomic magnetic moments are computed by integrating over a sphere, different magnetization densities within the sphere may in principle lead to identical moments. 

\begin{figure}
	\centering
    \includegraphics[width=8.5cm]{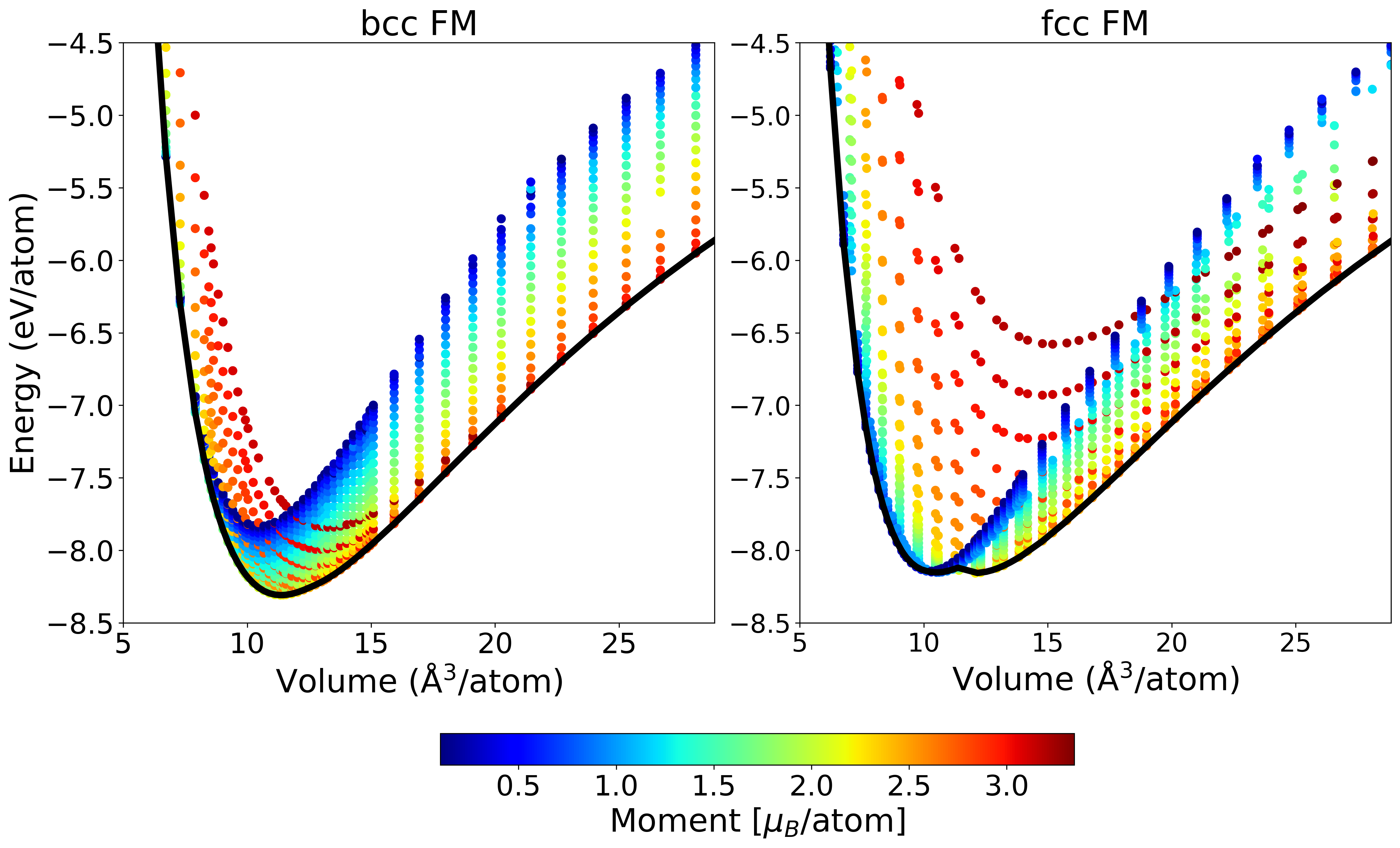}
	\caption{Constant magnetic moment energy-volume curves for FM bcc and fcc phases computed using constrained DFT. The black curve marks the ground state configurations without any applied constrain. The two minima for fcc correspond to the high- and low-spin magnetic configurations.}
	\label{fig:evdata}
\end{figure}

To generate the training dataset for magnetic ACE, we considered both conventional, unconstrained and collinear as well as constrained non-collinear spin polarized configurations. These configurations ranged from various spin spirals in ideal bcc cells to supercells with random orientations of the moments and perturbed atomic positions. For bulk phases along the Bain transformation path, we sampled the  magnitudes of the collinear magnetic moment over the whole physically reasonable range from 0 to $\sim$3 $\mu_B$/atom. The simultaneous sampling of both atomic and magnetic DOF enabled to generate a set of uniformly distributed configurations that are relevant for the properties of interest for a wide range of atomic densities as well as magnitudes and directions of the atomic magnetic moments. An example of data collected with this strategy is given in Fig.~\ref{fig:evdata} for the bcc and fcc ferromagnetic (FM) phases. Each data point corresponds to the energy of either structure at a given volume and a constrained value of the magnetic moment. The ground state configurations are marked by the black curve. While the bcc phase has only one minimum, corresponding to the ground state FM bcc phase, the fcc phase exhibits two minima corresponding to high-spin and low-spin configurations.

The constrained magnetic calculations required convergence of the energy and forces with respect to a constraining penalty term~\cite{PhysRevB.91.054420}. In some limited cases it was computationally prohibitive to achieve numerically small penalty contributions, mainly for configurations far from equilibrium such as highly distorted structures and defects (see SI for representative examples). Therefore, we excluded configurations for which the penalty energy was larger than $\approx$ 5 meV/atom as these would significantly increase the noise in the data and adversely affect the parameterization.

The resulting training dataset contained about 70 000 structures  in total that can be divided into several categories, each associated with a particular property of interest. The categories are listed in Table~\ref{table:1}, where we specify the number of configurations and the range of volumes and magnetic moment magnitudes that we considered. The free atom data, obtained from calculations of a bcc unit cell with lattice parameter equal to 12 $\angstrom$ at different magnitudes of the magnetic moment, is used to fit the first order contribution of the expansion that characterizes the asymptotic large volume limit for each structure with a given magnetic moment. Detailed information about the free atom reference is provided in the supplementary material.



\begin{table*}[t]
	\centering
			\caption{A list of categories associated with a given target property. To each category we provide the total number of structures, the volume range in percentage of the equilibrium volume $V_{0}$ of the corresponding phase, the range of sampled magnetic moments,  and the number of atoms in the simulation cell. In the case of both constrained and unconstrained supercell calculations, only the direction of the magnetic moments was fixed while their magnitude was self-consistently converged to the equilibrium value. Defects were calculated without any constrains regarding the atomic magnetic moments.}
		\label{table:1}
		\begin{tabular}{| l | c | c | c | c |} 
			\hline
			Property & Number of  & Volume & M range & Number of atoms \\  
			& structures & range ($\%V_{0}$) & ($\mu_{B}$) &  per cell \\ [0.5ex] 
			\hline
			Free atom & 15 & -- & 0.0-4.0 & 1 \\
			E-V curves & 10030 & $\pm$30 & 0.0-3.2 & 2,4 \\ 
			Elastic constants & 13919 & $\pm$5 & 1.5-2.8 & 2 \\
			Phonons & 2847 & $\pm$5 & 1.5-2.8 & 4-12 \\
			Supercells & 13762 & $\pm$5 & -- & 16,32,54 \\
			Transformation paths & 16805 & $\pm$20 & 0.0-3.2 &  2,4\\
			Spin rotations & 5559 & $\pm$30 & 0.0-3.2 & 2 \\ 
			Defects & 12233 & $\pm$20 & -- & 3-129 \\[1ex] 
			\hline
		\end{tabular}
\end{table*}

\subsection{Training procedure \label{sec:training}}

The fitting of the magnetic ACE potential for Fe was done following procedures that were established for the nonmagnetic ACE~\cite{lysogorskiy2021performant,bochkarev_PhysRevMaterials.6.013804}. 
A hierarchical basis extension was employed, starting from one-body contribution and adding gradually contributions with higher body orders. 
The first step of the fitting procedure consisted in the parameterization of the expansion coefficients for the first order contribution using the free magnetic atom data. 
This term can be reduced to a Ginzburg-Landau expansion $\sum_{n}A_{n}\textbf{m}_{i}^{2n}$, where a maximum number of three terms is commonly used~\cite{PhysRevB.86.054416,PhysRevB.55.14975,PhysRevB.81.184202,PhysRevLett.77.334,PhysRevB.102.014402}.  
In our parameterization, excellent agreement with the reference data ($\sim$2 meV/atom error) could be obtained using four terms ($n=4$). After the first order contribution was fixed, we fitted second order contributions. The ACE second order contributions are formally equivalent to a distance dependent Heisenberg Hamiltonian $\sum_{i>j}J_{ij}(r_{ij})\textbf{m}_{i}\cdot\textbf{m}_{j}$, its biquadratic correction $\sum_{i>j}B_{ij}(r_{ij})(\textbf{m}_{i}\cdot\textbf{m}_{j})^{2}$, and its bicubic term for $l'_{max}=1$, 2 and 3, respectively (see Methods).  In addition, a third-order magnetic contribution, analogous to a screened three-spin interaction $\sum_{ijk}K_{ijk}(\textbf{m}_{i}\cdot\textbf{m}_{j})(\textbf{m}_{j}\cdot\textbf{m}_{k})$, was also included. Angular contributions in higher order magnetic terms did not improve the fit significantly and were neglected, which reduced the number of basis functions significantly.   Radial and angular indices for the atomic contributions were then incremented following a hierarchical basis expansion scheme~\cite{bochkarev_PhysRevMaterials.6.013804}, where contributions with increasing body order were gradually added. The cutoff distance of the ACE was set to $4.5$~$\angstrom$. However, it can be extended if necessary ~\cite{PhysRevLett.107.017204}. Additional hyperparameters, relevant to magnetic DOF only, include the magnetic cutoff $m_{cut}$ set to 4 $\mu_{B}$, which defines the upper bound of the possible magnitude of atomic magnetic moments, and the upper bounds for magnetic radial functions and spherical harmonics $n'_{max}$ and $l'_{max}$ for each body order (see Methods for details).

The resulting model consists of 6519 parameters and its overall accuracy is equal to 8 meV/atom and 37 meV/$\angstrom$
for energies and forces, respectively. The main limiting factor in reducing the error further was numerical noise in the reference data that originated from the magnetic moment confinement procedure. In addition, another parameterization was constructed with a particular focus on defect properties (see supplementary material).

\subsection{Properties of the non-collinear magnetic ACE}

\begin{figure}[h]
	\centering
    \includegraphics[width=8.5cm]{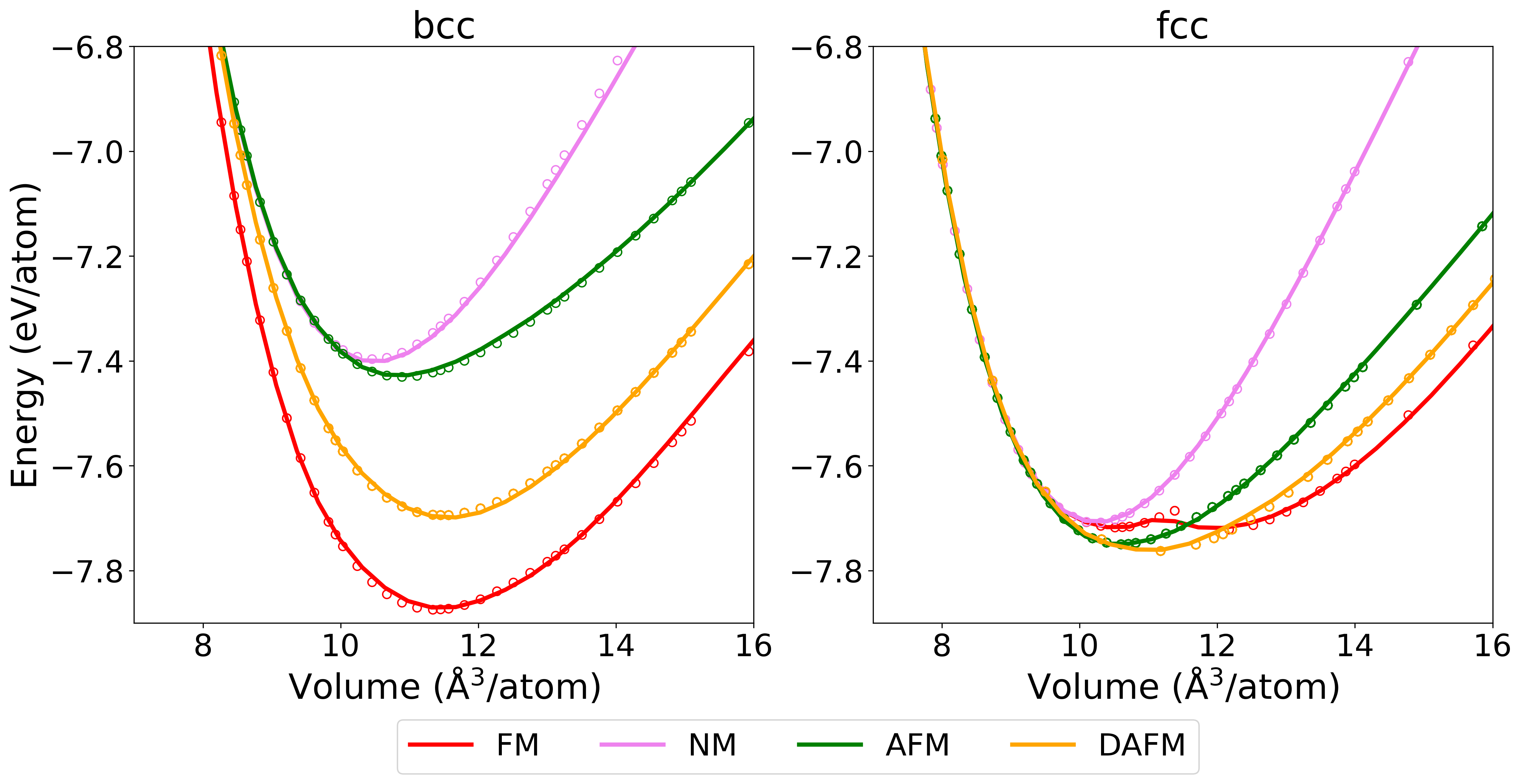}
	\caption{Volume-energy curves for both magnetic and non-magnetic structures of bcc (left) and fcc (right) with corresponding DFT data (small circles).}
	\label{fig:evmace1}
\end{figure}

We carried out a thorough validation of magnetic ACE against the reference DFT data and evaluated a broad range of properties of various bulk Fe phases that were not included explicitly in the training. The predicted volume-energy curves for the bcc and fcc magnetic and non-magnetic (NM) Fe phases are plotted in Fig.~\ref{fig:evmace1}, where the corresponding cohesive energies are given with respect to the non-magnetic free atom.  It is obvious that ACE predictions agree closely with the reference DFT data for all considered magnetic and non-magnetic phases, including the portion of the magnetic energy landscape where the NM to magnetic transitions take place. Moreover, our potential is able to distinguish  subtly different  magnetic states within one structure, such as  the low-spin and high-spin states of the FM fcc structure.

\begin{figure}[h]
	\centering
    \includegraphics[width=8.5cm]{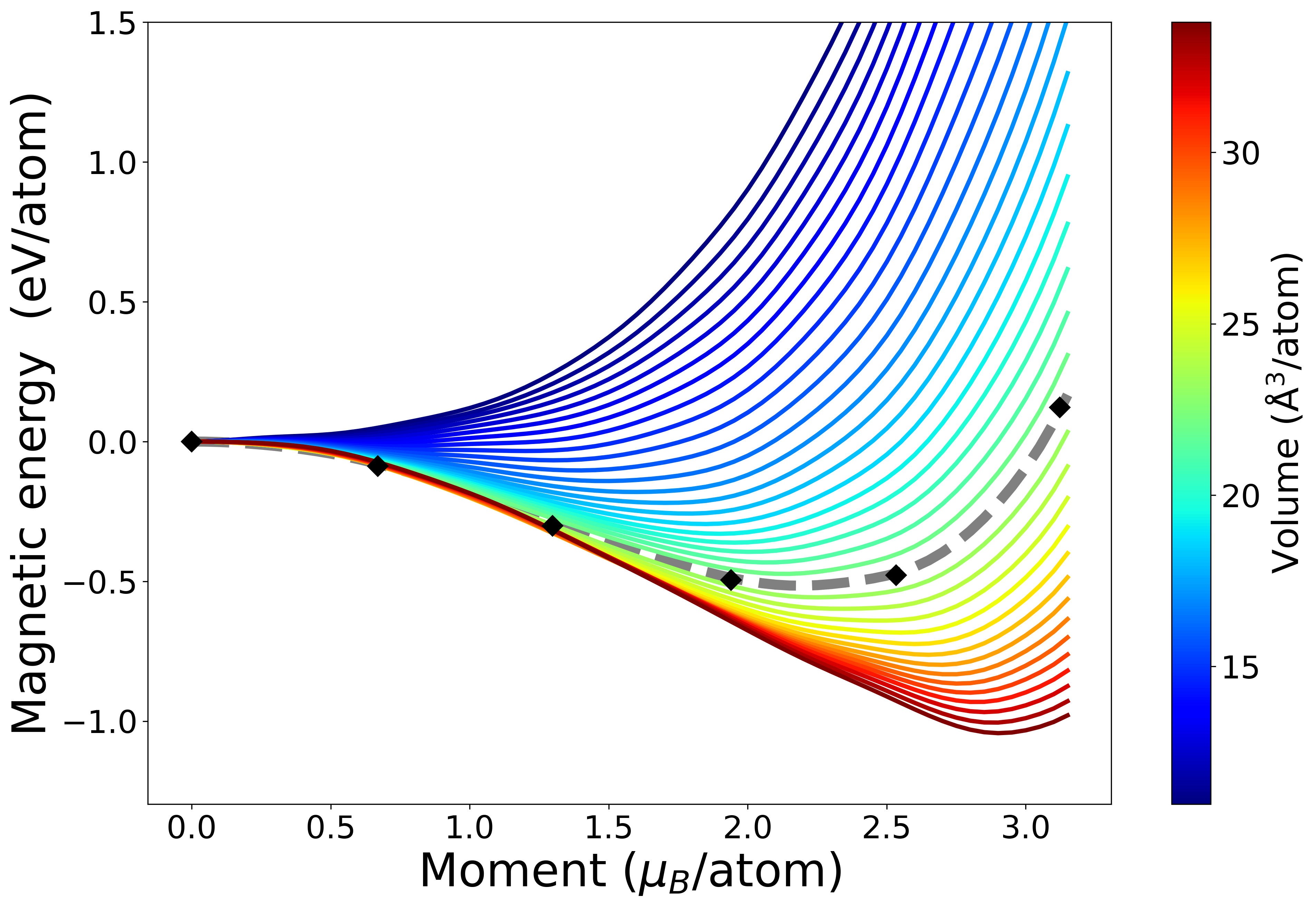}
	\caption{Magnetic energy vs magnetic moment magnitude at different volume for bcc FM. Dashed lines and black dots correspond to equilibrium volume ACE and reference DFT data, respectively.}
	\label{fig:embccFM}
\end{figure}

Variations of the magnetic energy as a function of magnetic moment are displayed for FM bcc in Fig.~\ref{fig:embccFM}, where each curve corresponds to a constant volume. As expected, these dependencies are positive and monotonic for small volumes (dark blue curves), while above a certain critical volume their behavior qualitatively changes to include a minimum at finite value of the magnetic moment in analogy to a Landau expansion. In the limit of large volumes (dark red curves), the magnetic energy approaches the free atom value. Graphs for other bcc and fcc structures are given in the supplementary material.

\begin{figure}
    \includegraphics[width=8.5cm]{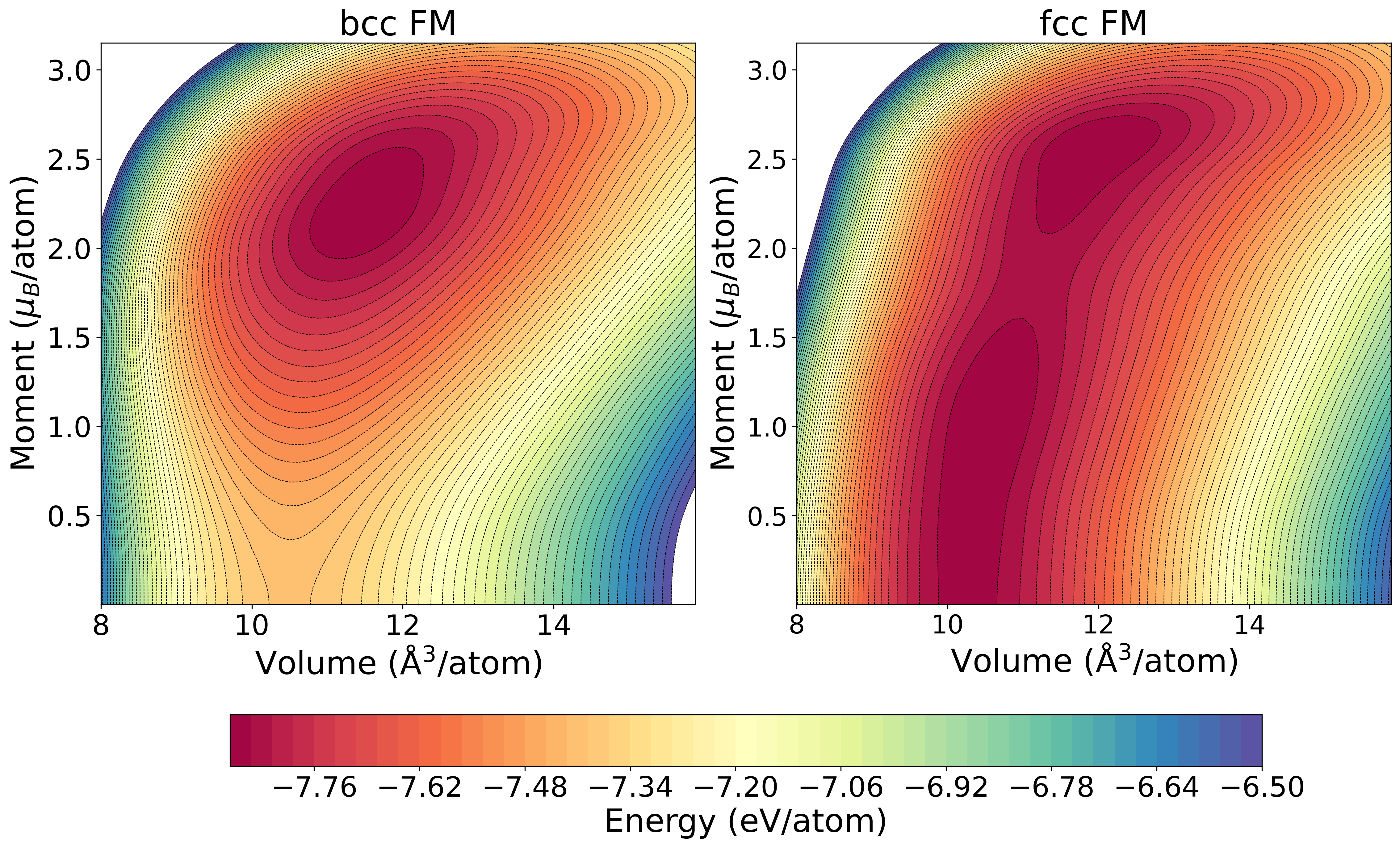}
    \caption{Contour plots for the bcc (left) and fcc (right) FM phases.}
    \label{fig:contouACE}
\end{figure}

Two contour plots of magnetic PES for FM bcc and fcc phases are shown in Fig.~\ref{fig:contouACE}. These plots demonstrate that ACE can capture simultaneously PES of different phases over a broad range of volumes and magnetic moments (from zero up to $\approx 3.2\, \mu_{B}$).  In agreement with DFT, the bcc phase has a single global minimum at the corresponding equilibrium volume and magnetic moment, while the FM fcc phase exhibits two local minima corresponding to low-spin and high-spin states.

Equilibrium properties of the most important bulk Fe phases are listed in Table~\ref{table:2}. Further properties, such as phonon spectra and magnetic moment variations, are presented in the supplementary material. As one can see, the equilibrium lattice parameters, magnetic moments, and elastic constants of the magnetic phases are in good agreement with the reference DFT values. Larger discrepancies exist for the non-magnetic phases since only very few of these configurations were included in the training dataset.

\begin{table*}[t]
	\setlength{\tabcolsep}{10pt} 
    \renewcommand{\arraystretch}{1.0}
\centering
\begin{tabular}{ l l c c c c c c}
\hline
 & & $a$ ($\angstrom$) & $M$ ($\mu_{B}$) & $C_{11}$ & $C_{12}$ & $C_{44}$ & $C'$ \\ 
\hline
\multirow{3}{*}{bcc} &  FM  & 2.83 (2.83) & 2.22 (2.20) & 302 (283) & 158 (145) & 95 (104) & 72 (69) \\ 
                    & AFM  & 2.80 (2.79) & 1.25 (1.35) &  46 (4)       & 252 (249) & 175 (139) & $-103$ ($-123$) \\
                    &  NM  & 2.74 (2.76) &             & $-44$ (87) & 187 (361) & 104 (180) & $-116$ ($-141$)\\
 & & & & & & & \\
\multirow{2}{*}{fcc} &  FM  & 3.61 (3.65) & 2.35 (2.63) & 236 (255) & 312 (133) & 80 (85) & $-38$ (61) \\
                    &  NM  & 3.46 (3.45) &             & 634 (414) & 249 (214) & 326 (240) & 193 (100) \\

\hline
\end{tabular}
    \caption{Equilibrium properties of bcc and fcc phases of Fe predicted by ACE and DFT (in brackets). The elastic constants are given in GPa; the elastic constant $C'=\frac{1}{2}(C_{11}-C_{12})$ characterizes the stability of the structures with respect to the Bain distortion. Data for FM fcc correspond to the high-spin state.}
    \label{table:2}
\end{table*}





\begin{figure}[h]
	\centering
	\includegraphics[width=8.5cm]{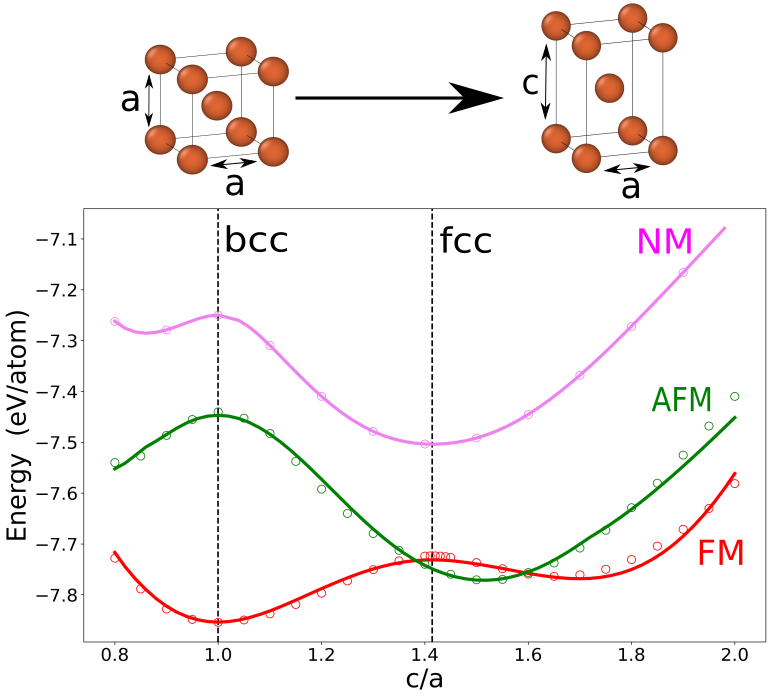}
	\caption{Bain transformation paths between the FM, AFM and NM bcc and fcc phases (ACE: lines, DFT: circles).}
	\label{fig:tp_mace}
\end{figure}

The Bain transformation path is closely related to the bcc-fcc phase transformation. In the case of Fe, the energetics of this transformation depends sensitively on the magnetic state of both phases~\cite{Herper_PhysRevB.60.3839, Okatov_PhysRevB.79.094111, Okatov_PhysRevB.84.214422, wang2018martensitic}. Variations of energy along the Bain path, computed at the FM bcc equilibrium volume for different magnetic phases of Fe are shown in Fig.~\ref{fig:tp_mace} for ACE and DFT. Unlike the ground state FM bcc phase, the AFM and NM bcc phases are not mechanically stable with respect to tetragonal distortion, as reflected by negative values of the $C'=\frac{1}{2}(C_{11}-C_{12})$ elastic constant (cf. Table~\ref{table:2}). For the fcc phase ($c/a=\sqrt{2}$), the energies of the FM and AFM magnetic states are almost identical, but both phases are unstable. The minimum energy AFM structure is a body-centered tetragonal phase with $c/a \approx 1.45$. The excellent agreement between ACE and DFT for the Bain path is due incorporating the coupling between the magnetic and lattice DOF correctly,  which is anomalously strong in Fe~\cite{Okatov_PhysRevB.84.214422}.

\begin{figure}[h]
	\centering
    \includegraphics[width=8.9cm]{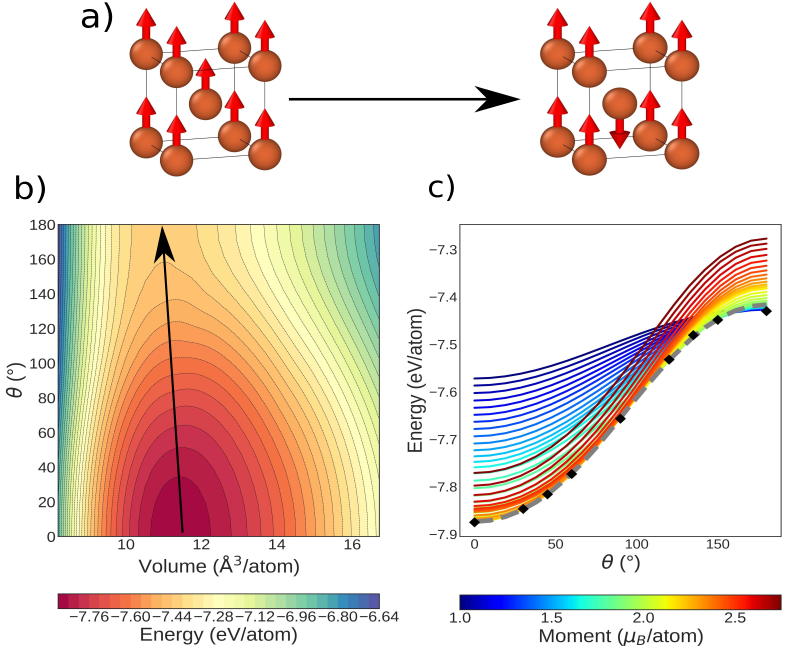}
	\caption{Analysis of the FM to AFM transformation in the bcc phase via rotation of the spin on the central atom: a) A schematic picture of the transformation. b) A contour plot of PES as a function of volume vs rotation angle. 
	c) FM to AFM spin rotation energy barriers at constant magnetic moment. The minimum energy path is marked by the grey dashed curve together with the DFT reference (black points).}
	\label{fig:spinrotmag_mace}
\end{figure}

The energy barrier for spin rotations depends sensitively on angular interactions between atomic magnetic moments and changes in moment magnitudes. Here we demonstrate that ACE captures the energetics of spin rotation between FM and AFM bcc phases. In Fig.~\ref{fig:spinrotmag_mace}, we show the energetics associated with the rotation of one magnetic moment in a two-atom bcc cell. As the moment on the central atom is rotated, the magnetic configuration gradually transforms from FM to AFM. The contour plot in Fig.~\ref{fig:spinrotmag_mace}(b) depicts PES as a function of volume and rotation angle. The black arrow marks the minimum energy path between the FM and AFM phases. While the equilibrium volumes of both phases are not very different, the magnetic moment of the AFM phase is significantly lower than that of the FM phase (cf. Table~\ref{table:2}). This is also correctly reproduced by ACE, as shown in Fig.~\ref{fig:spinrotmag_mace}(c), where we plot the rotation energy barriers evaluated at constant magnetic moments. The minimum energy path (dashed grey curve), corresponding to a reduction of the absolute value of magnetic moment from 2.22 $\mu_B$ in FM bcc to 1.25 $\mu_B$ in AFM bcc, is in excellent agreement with the DFT reference (black points).

\begin{figure}[h]
	\centering
    \includegraphics[width=9.0cm]{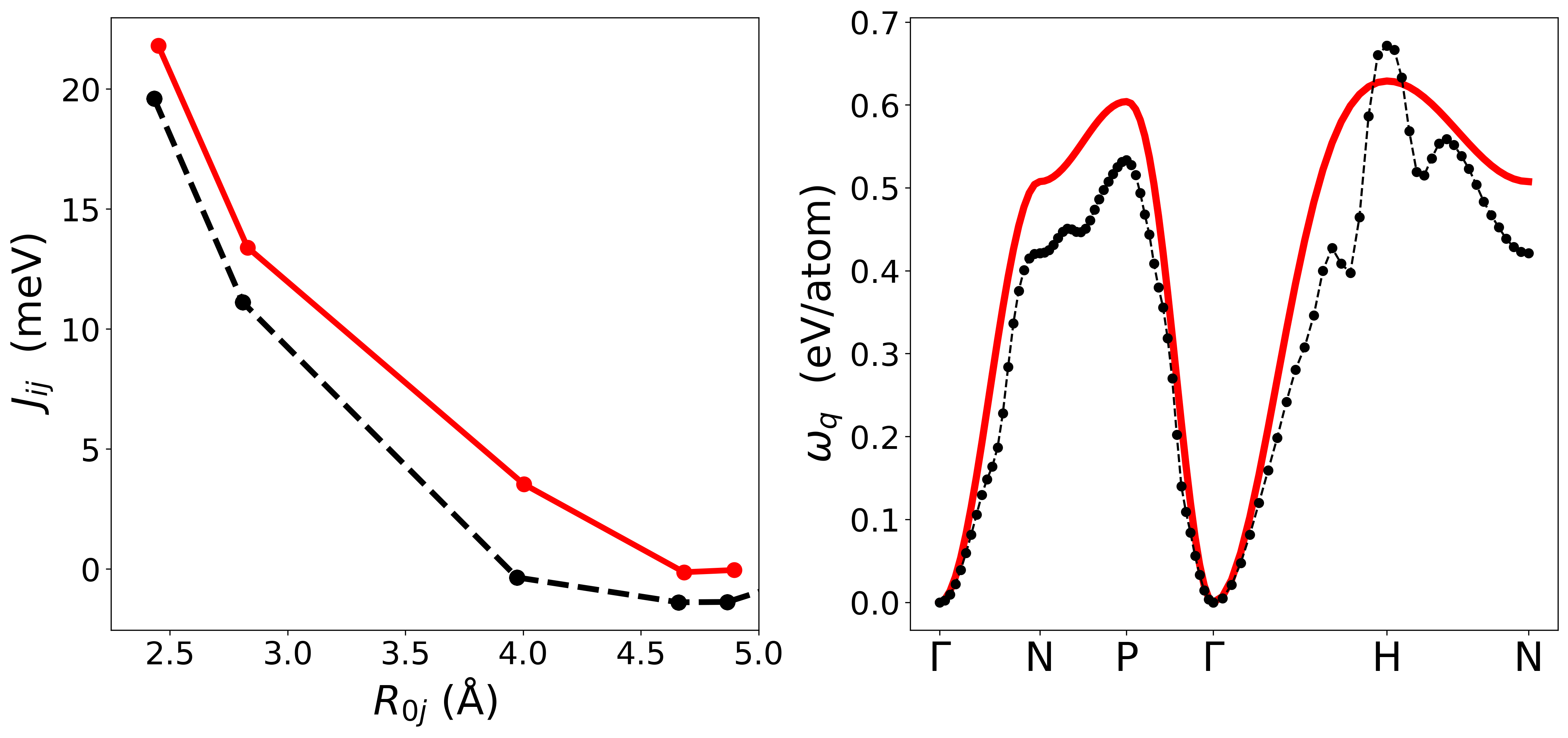}
	\caption{Exchange interactions (left) and adiabatic magnon spectra (right) predicted by ACE (red) in comparison with SPRKKR calculations (black).}
	\label{fig:magnon_mace}
\end{figure}

The energy of magnetic moment orientations that deviate only slightly from the collinear alignment can be described by lowest order contributions only, i.e., a bilinear Heisenberg model.  From the distance dependent exchange interactions $J_{ij}$ in the bilinear Heisenberg model, the magnon spectrum can be obtained in adiabatic approximation as
\begin{equation}
E_i\left(\textbf{q}\right)=\sum_{j}J_{ij}\left[1-\cos\left(\textbf{q}\cdot\textbf{R}_{ij}\right)\right].
\end{equation}
We determined the exchange interactions for different coordination shells following the real space approach by Liechtenstein et al.~\cite{liechtenstein1984exchange, liechtenstein1987local,
liechtenstein1986lsdf}, where infinitesimal perturbations to the directions of two neighboring magnetic moments are applied.  Calculating the energy $\delta E_{ij}$ for rotating two spins at atomic sites $i$ and $j$ by opposite infinitesimal angles $\pm\theta/2$ and comparing to the energy for rotating the two spins individually, $\delta E_{i}$ and  $\delta E_{j}$, results in
\begin{equation}
\delta E_{ij} - (\delta E_{i} + \delta E_{j}) = J_{ij}\left(1-\cos\theta\right)\sim\frac{1}{2}J_{ij}\theta^{2}.
\end{equation}
The distance dependent exchange interactions are then obtained by fitting $\delta E_{ij} - (\delta E_{i} + \delta E_{j})$ with respect to the tilting angle for consecutive coordination shells in a large supercell.  The resulting adiabatic magnon spectrum is shown in Fig.~\ref{fig:magnon_mace} with reference data obtained using the spin-polarized relativistic Korringa-Kohn-Rostoker (SPRKKR)~\cite{sprkkr} framework.  Small discrepancies between the ACE and SPRKKR results visible for some high frequencies in the magnon spectrum can be attributed to the long range part of the magnetic interactions neglected in the present ACE parameterization. Nevertheless, the overall agreement is good, indicating the ability of our parameterization to describe spin spirals with different frequencies.

\subsection{Finite temperature and defects}

\begin{figure}[h]
	\centering
    \includegraphics[width=8.0cm]{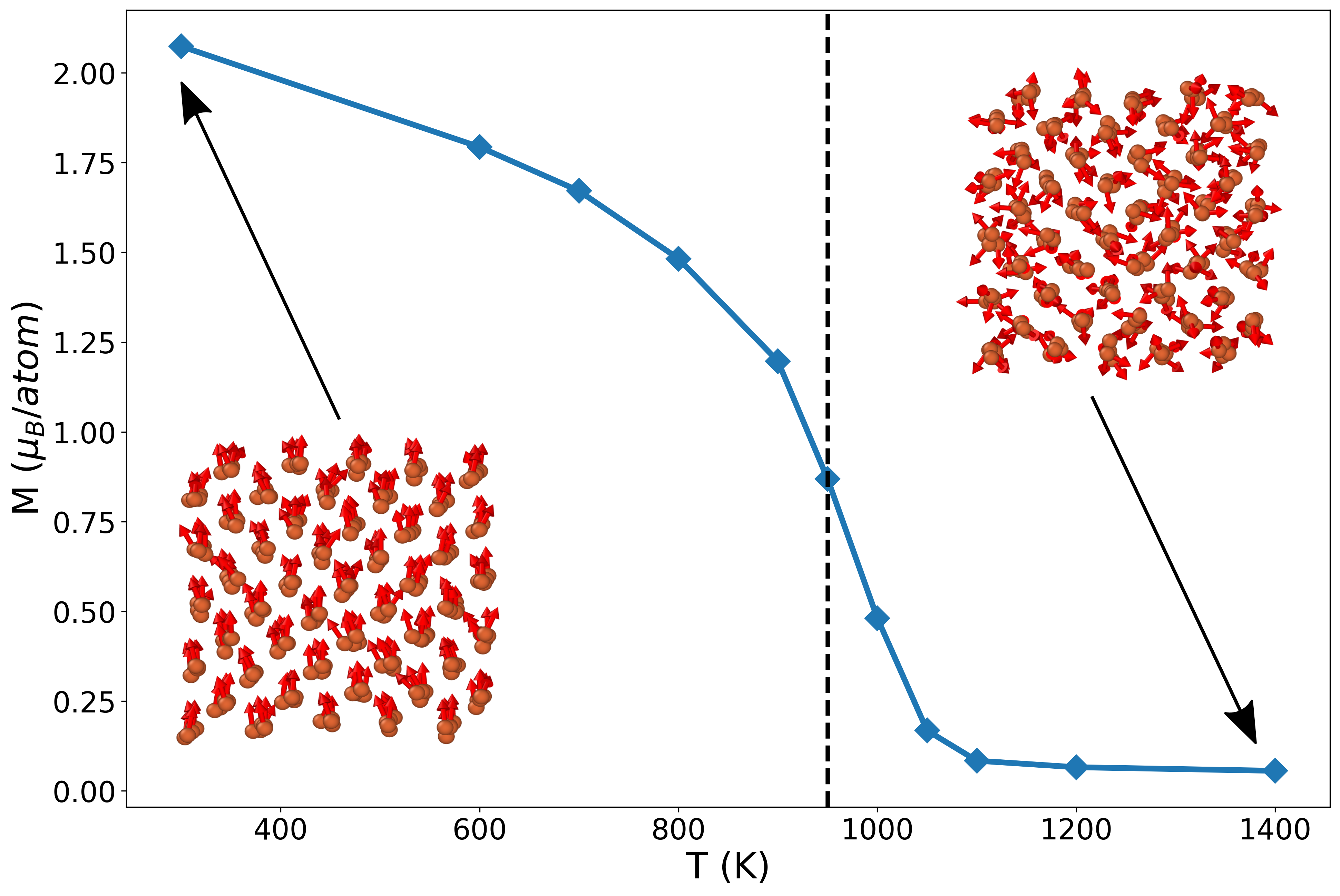}
	\caption{Magnetization vs temperature. The vertical dashed line indicates the estimated Curie temperature $T_{C}$. The experimental value of $T_{C}$ is 1043 K. Insets show snapshots of parts of the simulation cell for better visualization.}
	\label{fig:avg_mag}
\end{figure}

The magnetic ACE can be applied in large-scale finite temperature simulations to investigate properties that depend on both spin and lattice DOF. The ACE prediction of the FM to paramagnetic (PM) phase transition in bcc Fe is presented in Fig.~\ref{fig:avg_mag}. In a simulation with 3456 atoms, we employed coupled molecular dynamics (MD) - Monte Carlo (MC) sampling~\cite{PhysRevLett.121.125902}, where the atoms follow Langevin dynamics while MC is employed for updating the directions of the atomic magnetic moments at constant magnitudes. A direct simulation of the dynamics of the combined atomic and magnetic system is difficult due to the lack of numerically stable and efficient symplectic integrators \cite{Wang_PhysRevB.99.094402}. The predicted Curie temperature $T_{C} \approx 950$~K is somewhat underestimated in comparison with the experimental value of 1043~K. The variation of magnetization with temperature shown in Fig.~\ref{fig:avg_mag} is consistent with previous theoretical studies~\cite{PhysRevB.96.094418, PhysRevB.78.024434, PhysRevB.86.054416, PhysRevB.78.033102}. 

To demonstrate that ACE is able to capture properties of crystal defects, we also included several defect configurations in the DFT training data. However, as discussed in Sec.~\ref{sec:DFT}, it is often not possible to reach sufficiently small penalty energies in the constrained DFT calculations for such distorted configurations. Therefore, we needed to resort in many cases to unconstrained spin-polarized calculations only, which limited the sampling of the magnetic PES for defects.

Here we present results for three types of defects - a monovacancy, generalized stacking faults, and a screw dislocation. For most defects, the Heisenberg model is insufficient and it is necessary to include higher-order terms in the magnetic Hamiltonian~\cite{chapman2020effect}. In addition, an accurate reproduction of defect properties can be achieved only if the coupling between spin and lattice excitations is taken into account. 

\begin{figure}[h]
	\centering
    \includegraphics[width=9.0cm]{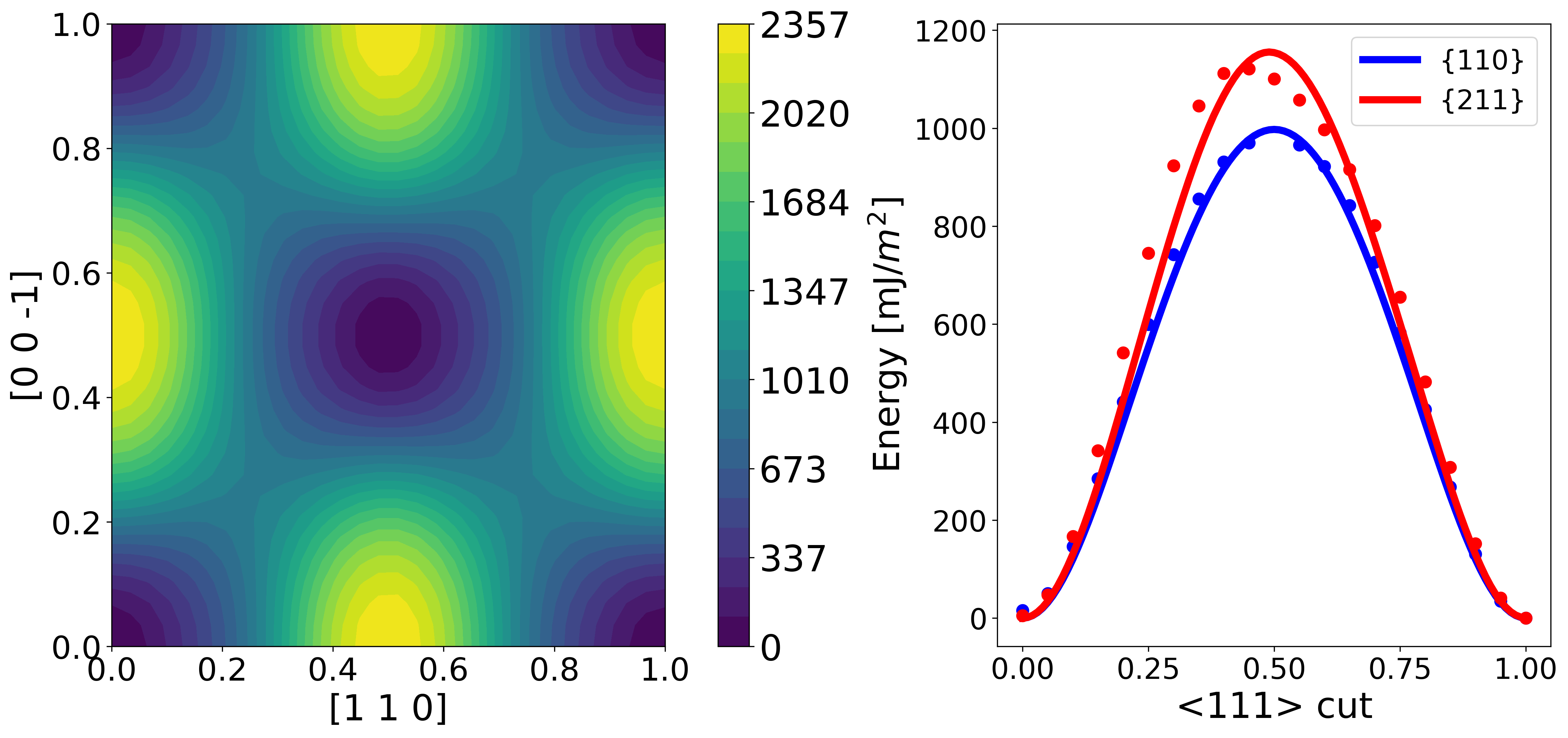}
	\caption{Predicted $\gamma$-surface for the $\{110\}$ crystallographic plane (left) and cuts along the $\langle 111 \rangle$ direction for the $\{110\}$ and $\{211\}$ $\gamma$-surfaces (ACE: lines, DFT: dots) (right).}
	\label{fig:cut_gamma}
\end{figure}

\begin{figure}[h]
	\centering
	\includegraphics[width=6.0cm]{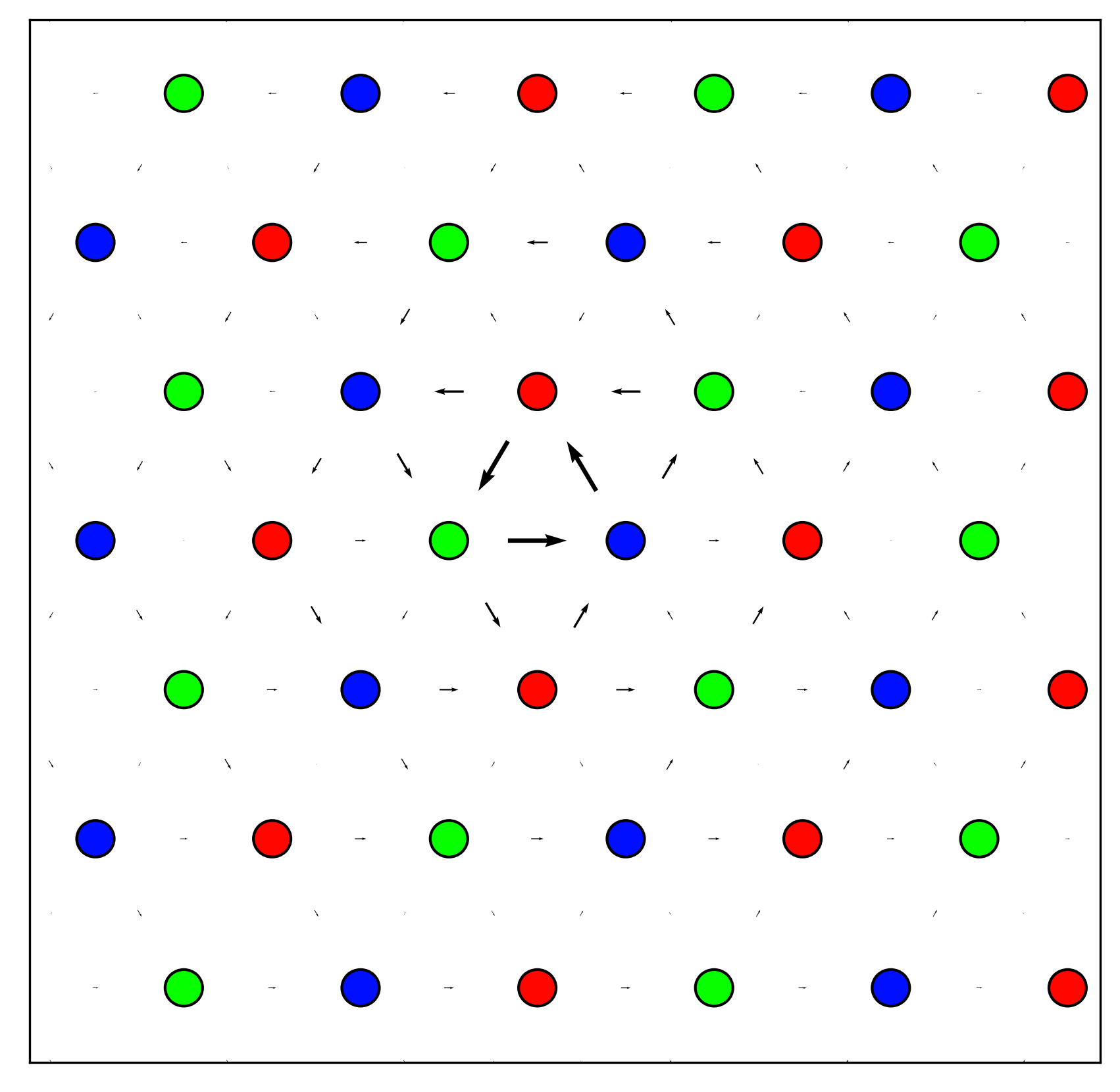}
	\caption{Differential displacement map of the $\frac{1}{2} \langle 111 \rangle$ screw dislocation predicted by the magnetic ACE potential.}
	\label{fig:screw}
\end{figure}

The monovacancy formation and migration energies of 2.57 eV and 0.65 eV, respectively, agree well with the reference DFT data (equal to 2.17 and 0.67 eV, respectively).  The generalized stacking fault energy surface, the $\gamma$-surface, for the $\{110\}$ plane is shown in Fig.~\ref{fig:cut_gamma}. The figure further shows cuts along the $\langle 111 \rangle$ direction on both the $\{110\}$ and $\{211\}$ planes that are related to atomic structures of $\frac{1}{2} \langle 111 \rangle$ screw dislocations. The ACE predictions for both cuts are in excellent agreement with the DFT reference. ACE also predicts the core structure of the $\frac{1}{2} \langle 111 \rangle$ screw dislocation, which governs the low temperature plasticity of Fe, in quantitative agreement with DFT reference \cite{ventelon2007core, dezerald2014ab} as demonstrated in Fig.~\ref{fig:screw}.

To examine the ability of ACE to reproduce properties of other defects, we constructed a smaller training dataset containing also surfaces and interstitials and generated another ACE parameterization. As shown in the supplementary material, ACE can reproduce properties of these defects as well.

\section{Discussion}

By incorporating magnetic DOF in the form of atomic magnetic moment vectors into ACE, we demonstrated that constrained non-collinear DFT reference data can be reproduced with excellent accuracy and transferability, exceeding those of existing magnetic ML interatomic potentials. We constructed a non-collinear ACE parametrization for Fe and validated it for a wide range of properties, including volume-energy curves, elastic moduli, phonon spectra, Bain transformation paths, spin rotations and magnon spectra, and point and extended defects. These tests showed that magnetic ACE is not only able to capture large structural and magnetic variations but also resolves subtle spin fluctuations  that are crucial for a correct reproduction of phase transitions and thermodynamic properties. To this end it is necessary to include multi-atom multi-spin interactions that are missing in simple models with pairwise couplings between atoms and/or magnetic moments.  In iron 
magnetic angular contributions of body order four and higher are numerically small and can be neglected.

The magnetic ACE was parameterized from DFT reference data that was generated by constraining both the magnitude and direction of the atomic magnetic moments. For configurations with defects or significant atomic displacements, it was often difficult to achieve self-consistency. Furthermore, as the constraints were implemented by integrating magnetization over spheres about atoms, various intra-atomic magnetization distributions could result in the same atomic magnetic moment. This non-uniqueness effectively led to noise in the DFT reference data that ultimately limited the accuracy of our parameterization. Therefore, there is a strong incentive to implement more advanced constraints in DFT that would help to increase the accuracy of magnetic ACE as well as other magnetic ML approaches.

The numerical efficiency of ACE enables to carry out large-scale molecular and spin dynamics simulations to study the dynamics of combined magnetic and structural phase transitions.  Nevertheless, to predict the magnetic transition in bcc iron, we employed MD simulations for the atomic positions and MC sampling to vary the atomic magnetic moments. One of the reasons is that there is a lack of classical or semi-classical equations of motion and corresponding, numerically robust integrators applicable to combined atomic and spin dynamics in systems with multi-atom multi-spin interactions and including changes of magnitudes of magnetic moments~\cite{Wang_PhysRevB.99.094402}.

The ACE for iron can be extend directly to multicomponent systems, such as technologically important magnetic alloys and carbides. While this is straightforward from a formal point of view, the generation of accurate and comprehensive DFT reference data for magnetic multicomponent materials is challenging. Here efficient sampling based on D-optimality active learning~\cite{lysogorskiy2023active} extended to include magnetic DOF will help to reduce the number of required DFT reference calculations.

\section{Methods}

We provide a summary of the magnetic ACE formalism together with aspects of its implementation. Further explanations on implementation and workflow are available in the supplementary material. We also provide computational details of the DFT calculations and the combined MD-MC simulations that were employed for the calculation of the FM-PM transition. 

\subsection{Energies, forces and magnetic gradients}

We define state variables $\sigma_{ji}$  of atom $j$ neighboring atom $i$ in terms of interatomic distances vectors $\textbf{r}_{ji}$, chemical species $\mu_{j}$, , magnetic moments $\textbf{m}_{j}$, etc. as
\begin{equation}
\sigma_{ji}=\left(\mu_{j},\textbf{r}_{ji},\textbf{m}_{j}\right) \, ,
\end{equation}
with $\sigma_{ii}=\left(\mu_{i},\textbf{m}_{i}\right)$. A neighbor density on atom $i$ including atomic and magnetic contributions can then be written as
\begin{equation}\label{eq:macedens}
\varrho_{i}\left(\sigma\right)=\sum_{j}\delta\left(\sigma-\sigma_{ji}\right) \,.
\end{equation}

Magnetic contributions also enter the single bond basis functions,
\begin{equation}
\phi_{v}\left(\sigma_{ji}\right)=\delta_{k}\left(\mu_{j}\right)R_{nl}^{\mu_{j}\mu_{i}}\left(r_{ji}\right)Y_{l}^{m}\left(\hat{\pmb{r}}_{ji}\right)M_{n'l'}^{\mu_{j}\mu_{i}}\left(m_{j}\right)Y_{l'}^{m'}\left(\hat{\pmb{m}}_{j}\right)
\end{equation}
where $v = (knlmn'l'm')$ and the primed indices are used to label basis functions that depend on magnetic contributions, and $\phi_{v}\left(\sigma_{ii}\right)=\delta_{k}\left(\mu_{i}\right)M_{n'l'}^{\mu_{i}\mu_{i}}\left(m_{i}\right)Y_{l'}^{m'}\left(\hat{\pmb{m}}_{i}\right)$. 

The projection of the density in Eq.~(\ref{eq:macedens}) on the corresponding single atom basis functions leads to the atomic basis $A_{iv}$ and $A_{iv}^{(0)}$
\begin{equation}\label{eq:aiv}
A_{iv}=\Braket{\rho_{i}|\phi_{v}}=\sum_{j\neq i}\phi_{v}\left(\sigma_{ji}\right)
\end{equation}
and
\begin{equation}\label{eq:aiv0}
A_{iv}^{(0)}=\Braket{\rho_{i}^{(0)}|\phi_{v}}=\phi_{v}\left(\sigma_{ii}\right).
\end{equation}
From the two atomic bases the tensor product basis is formed
\begin{equation}\label{eq:ncorr}
\textbf{A}_{i\textbf{v}}=A_{iv_{0}}^{(0)}\prod_{t=1}^{N}A_{iv_{t}},
\end{equation}
and symmetrized to ensure invariance with respect to rotation and inversion, leading to equivariant basis functions
\begin{equation}\label{eq:fgeneral}
\textbf{B}_{i}=\mathcal{C}\cdot\textbf{A}_{i},
\end{equation}
where $\mathcal{C}$ is a sparse matrix of products of the Clebsch-Gordan coefficients of the atomic and magnetic systems. The coupling tree, used to form possible tuples $\textbf{v}$ (see SI for an example), can be simplified assuming that spin-orbit coupling can be neglected. This is typically an excellent approximation as the spin-orbit coupling energy is on the order of a few $\mu$eV for iron bulk systems. Then the atomic and magnetic systems can be completely decoupled and the total angular momenta of the atomic and magnetic channels couple to zero individually, leading to a significant reduction in the number of basis functions (see SI for a detailed explanation). A further reduction of the allowed combinations of atomic and magnetic indices can be obtained by requiring inversion invariance for both atomic and magnetic spaces by restricting the sum of the corresponding angular momenta to even numbers. 

We represent the energy for atom $i$ including atomic and magnetic contributions as a linear expansion
\begin{equation}\label{eq:expt}
\varepsilon_{i}=\textbf{c}^T \textbf{B}_{i} ,
\end{equation}
where $\textbf{c}$ is the vector of the expansion coefficients.

The energy can be rewritten in terms of the $\tilde{\textbf{c}}$ basis introduced in~\cite{Drautz_PhysRevB.102.024104,lysogorskiy2021performant} as
\begin{equation}\label{eq:condes}
\varepsilon_{i} = \textbf{c}^{T}\textbf{B}_{i} = \textbf{c}^{T}\mathcal{C}\textbf{A}_{i} = \tilde{\textbf{c}}^{T} \textbf{A}_{i}.
\end{equation}
This expansion was used to fit the DFT energies and forces.  In order that the expression reduces to the non-magnetic ACE when the magnetic moments are zero, the first order equivariant basis was taken as $A_{i\mu_{i}n'}^{(0)}\left(\textbf{m}=\textbf{0}\right)$=1 by our choice of magnetic radial functions (see the following Sec. \ref{sec:radial}).

Expressions for forces and magnetic gradients are obtained by taking the derivative of the energy with respect to atomic positions and magnetic moments, respectively, and are written in a compact notation as
\begin{equation}
\textbf{F}_{k}=\sum_{i}\left(\textbf{f}_{ik}-\textbf{f}_{ki}\right),
\end{equation}
and
\begin{equation}\label{eq:maggrad}
\textbf{T}_{k}=\sum_{i}\textbf{t}_{ki}+\textbf{t}_{k}.
\end{equation}
The pairwise atomic forces $\textbf{f}_{ki}$ are given by
\begin{equation}\label{eq:pseudof}
\textbf{f}_{ki}= \sum_{nlmn'l'm'}\omega_{i\mu_{k}nlmn'l'm'}\nabla_{\textbf{r}_{ki}}\phi_{\mu_{k}\mu_{i}nlmn'l'm'}
\end{equation}
and magnetic forces $\textbf{t}_{k}$ and $\textbf{t}_{ki}$ by
\begin{equation}\label{eq:tk0}
\textbf{t}_{k}=\sum_{n'l'm'}\omega_{k\mu_{k}n'l'm'}^{(0)}\nabla_{\textbf{m}_{k}}A_{k\mu_{k}n'l'm'}^{(0)}
\end{equation}
and
\begin{equation}\label{eq:tki}
\textbf{t}_{ki}= \sum_{nlmn'l'm'}\omega_{i\mu_{k}nlmn'l'm'}\nabla_{\textbf{m}_{k}}\phi_{\mu_{k}\mu_{i}nlmn'l'm'}.
\end{equation}
The calculation of the adjoints $\omega_{i\mu_{i}nlmn'l'm'}$ and $\omega_{i\mu_{i}n'l'm'}^{(0)}$ can be further decomposed to the evaluation of two distinct terms,
\begin{equation}\label{eq:adj}
\begin{split}
\omega_{i\mu_{i}nlmn'l'm'}&=\sum_{N=1}\sum_{\bm{\mu}\textbf{n}\textbf{l}\textbf{m}\textbf{n}'\textbf{l}'\textbf{m}'}\Theta^{(N)}_{\mu_{i}\bm{\mu}\textbf{n}\textbf{l}\textbf{n}'\textbf{l}'}\\ &\times A_{i\mu_{i} n_{0}'l_{0}'m_{0}'}^{(0)}\sum_{s=1}^{N}dA^{(s)}_{i\bm{\mu}\textbf{n}\textbf{l}\textbf{m}\textbf{n}'\textbf{l}'\textbf{m}'}
\end{split}
\end{equation}
where
\begin{equation}
\Theta^{(N)}_{\mu_{i}\bm{\mu}\textbf{n}\textbf{l}\textbf{n}'\textbf{l}'}=\tilde{c}_{\mu_{i}\bm{\mu}\textbf{n}\textbf{l}\textbf{n}'\textbf{l}'}^{(N)}
\end{equation}
and
\begin{equation}\label{eq:dbs}
\begin{split}
dA^{(s)}_{i\bm{\mu}\textbf{n}\textbf{l}\textbf{m}\textbf{n}'\textbf{l}'\textbf{m}'}&=\delta_{\mu\mu_{s}}\delta_{nn_{s}}\delta_{ll_{s}}\delta_{mm_{s}}\delta_{n'n_{s}}\delta_{l'l_{s}}\delta_{m'm_{s}}\\
& \times \prod_{k\neq s}A_{i\mu_{k}n_{k}l_{k}m_{k}n_{k}'l_{k}'m_{k}'}\,.
\end{split}
\end{equation}
The adjoint $\omega_{i\mu_{i}n'l'm'}^{(0)}$ does not contain the onsite basis contribution and is simply given by
\begin{equation}\label{eq:adj0}
\omega_{i\mu_{i}n'l'm'}^{(0)}=\sum_{N=0}\sum_{\bm{\mu}\textbf{n}\textbf{l}\textbf{m}\textbf{n}'\textbf{l}'\textbf{m}'} \Theta^{(N)}_{\mu_{i}\bm{\mu}\textbf{n}\textbf{l}\textbf{n}'\textbf{l}'} dA^{(0)}_{i\bm{\mu}\textbf{n}\textbf{l}\textbf{m}\textbf{n}'\textbf{l}'\textbf{m}'}
\end{equation}
with
\begin{equation}\label{eq:db0}
dA^{(0)}_{i\bm{\mu}\textbf{n}\textbf{l}\textbf{m}\textbf{n}'\textbf{l}'\textbf{m}'}=\prod_{s=1}^{N}A_{i\mu_{s}n_{s}l_{s}m_{s}n_{s}'l_{s}'m_{s}'}.
\end{equation}
The summation over $N$ in Eq.~(\ref{eq:adj0}) starts from zero because even a single atom contributes to the total magnetic gradient.

\subsection{Magnetic radial functions \label{sec:radial}}
The magnetic radial functions $M_{n'l'}^{\mu_{j}\mu_{i}}$ used in this work exhibit a different functional form to their atomic counterparts that are given in terms of Chebyshev polynomials~\cite{Drautz_PhysRevB.99.014104, bochkarev_PhysRevMaterials.6.013804}). In particular, one has to ensure that the energy is invariant under time reversal symmetry, i.e., $\textbf{m}_{i}\rightarrow -\textbf{m}_{i}$ for every $i$. For these reasons, we chose a linear combination of Chebyshev polynomials $T_{k}$ as
\begin{equation}\label{eq:rad_mag}
M_{n'l'}^{\mu_{j}\mu_{i}}\left(m\right)=\sum_{k'}c_{n'l'k'}^{\mu_{j}\mu_{i}}g_{k'}^{\mu_{j}\mu_{i}}\left(m\right),
\end{equation}
with
\begin{equation}
g_{k'}^{\mu_{j}\mu_{i}}\left(m\right)=T_{k'}\left(x(m)\right).
\end{equation}
The scaled distance $x$ guarantees the invariance under time reversal symmetry
\begin{equation}
x\left(m\right)=1-2\left(\frac{m}{m_{cut}}\right)^{2},
\end{equation}
where $m_{cut}$ is the cutoff for the magnetic moment magnitude. The expansion coefficients $c_{n'l'k'}^{\mu_{j}\mu_{i}}$ for both magnetic and atomic radial functions are adjusted during the fitting procedure.  

\subsection{DFT calculations}

All our reference DFT calculations were performed using the non-collinear and collinear versions of VASP 5.4.1~\cite{kresse1993ab,kresse1994ab,kresse1996efficiency,PhysRevB.54.11169} and the projector augment wave (PAW) method~\cite{blochl1994projector}. 
The constrained local moment approach~\cite{PhysRevB.91.054420} was employed to constrain either both size and direction or just the direction of the atomic magnetic moments. 
The exchange-correlation energy was represented using the Perdew-Burke-Ernzerhof (PBE) generalized gradient approximation (GGA) method~\cite{perdew1996generalized}.  
We carried out carefully converged calculations with tight settings of the principal parameters in order to obtain accurate results for the energy, forces and magnetic moments. Specifically, the kinetic energy cutoff was set to 500\,eV, the convergence threshold for the energy to 10$^{-5}$\,eV and the k-mesh density to 0.18\,$\angstrom^{-1}$. The integration radius for the atomic magnetic moments (VASP parameter RWIGS) was kept constant at the value of the Fe PAW (1.302\,$\angstrom$). 
See SI for the convergence of magnetic moment magnitude with respect to the integration radius and for a discussion on the convergence of the penalty energy in the constrained local moment method.

\subsection{MD-MC calculations}

The MD-MC simulations of the FM-PM transition in bcc Fe consisted of alternating MD and MC steps.  The MD simulations were performed using Langevin dynamics (from ASE~\cite{larsen2017atomic} package) with a time step of 1 fs. The MC sampling included uniform spin rotations on a unit sphere. The simulation supercell had dimensions $12\times 12\times 12$ of a bcc cell and contained 3456 atoms. The dimensions of the supercell were kept fixed at all temperatures so that the effect of thermal expansion was neglected. At each temperature, we carried out about $10^{7}$ steps, with the initial $10\%$ used for equilibration.

\section{Data availability}
The authors declare that all data supporting the findings of this study are available from MR on reasonable request.

\section{Code availability}
The magnetic ACE presented in the manuscript was evaluated and parameterized with a magnetic version of the PACEmaker code \cite{bochkarev_PhysRevMaterials.6.013804}. The magnetic ACE implementation is not yet distributed with PACEmaker, but we plan to make it available in a future release.

\section{Author contributions} 
MR implemented the magnetic ACE by extending non-magnetic ACE implementations in Fortran and in Tensorflow. MR produced the DFT reference data with support from MM and fitted the magnetic ACE. MR carried out all calculations presented in the manuscript and drafted the manuscript. MM and RD supervised the work, advised on theory and edited the manuscript. AB, YL and RD supported the implementation of magnetic ACE and advised on numerical aspects of the implementation.

\section{Competing Interests}
The authors declare no competing interests.

\section{Corresponding Author}
Correspondence should be addressed to MR and RD.

\begin{acknowledgments}
M.R. acknowledges support from the International Max Planck Research School for Interface Controlled Materials for Energy Conversion (IMPRS-SurMat). This work was in part supported by the German Science Foundation (DFG), projects 405621081 and 405621217. We acknowledge computational time from the Interdisciplinary Centre for Advanced Materials Simulation (ICAMS) and the Center for Interface Dominated High Performance Materials (ZGH), both at Ruhr University Bochum. 

\end{acknowledgments}

\newpage


\bibliographystyle{unsrt}

\newpage
\widetext
\clearpage
\begin{center}

\textbf{\large Magnetic Atomic Cluster Expansion and application to Iron: \\ Supplemental Materials}
\linebreak
\linebreak
\large{Matteo Rinaldi, Matous Mrovec, Anton Bochkarev, Yury Lysogorskiy, Ralf Drautz}
\end{center}
\setcounter{equation}{0}
\setcounter{figure}{0}
\setcounter{table}{0}
\setcounter{page}{1}
\setcounter{section}{0}
\renewcommand{\thesection}{\Roman{section}}
\makeatletter
\renewcommand{\theequation}{S\arabic{equation}}
\renewcommand{\thefigure}{S\arabic{figure}}
\renewcommand{\bibnumfmt}[1]{[S#1]}
\renewcommand{\citenumfont}[1]{S#1}

\section{Example of coupling tree}
\begin{figure}[h]
	\centering
	\includegraphics[width=10.0cm]{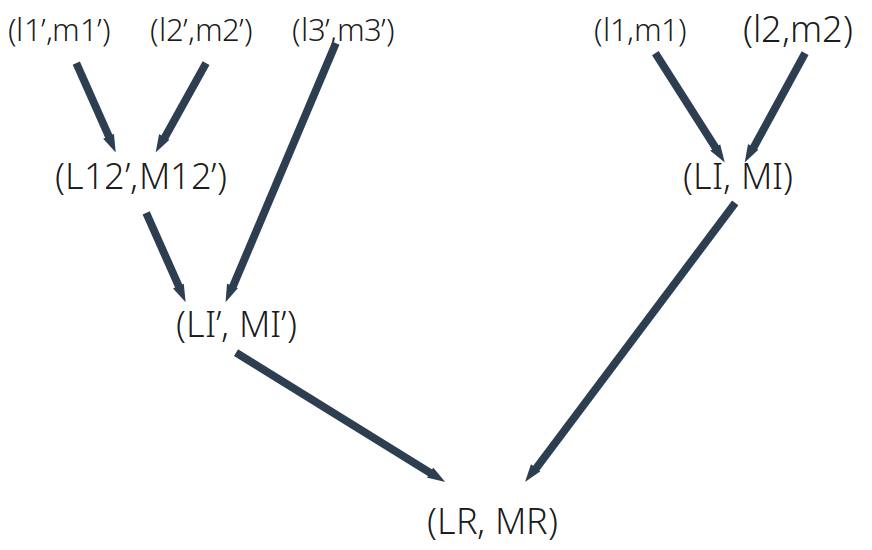}
	\caption{Coupling tree for third order contributions.}
	\label{fig:clbsch}
\end{figure}

An example of coupling tree is shown in Fig.~\ref{fig:clbsch} for the third order contribution. In terms of products of Clebsch-Gordan coefficients this reads,
\begin{equation}
\mathcal{C}_{l_{1}^{\prime}l_{2}^{\prime}L_{12}^{\prime}}^{m_{1}^{\prime}m_{2}^{\prime}M_{12}^{\prime}}\mathcal{C}_{L_{12}^{\prime}l_{3}^{\prime}L_{I}^{\prime}}^{M_{12}^{\prime}m_{3}^{\prime}M_{I}^{\prime}}
\mathcal{C}_{l_{1}l_{2}L_{I}}^{m_{1}m_{2}M_{I}}\mathcal{C}_{L_{I}^{\prime}L_{I}L_{R}}^{M_{I}^{\prime}M_{I}M_{R}}
\end{equation}
Here, the total angular momenta for the atomic and magnetic channels are denoted as $L_{I}$ and $L_{I}'$, respectively. They are coupled together to the total angular momentum $L_{R}$ of the mixed system, which is zero when expanding the energy, resulting in the identity $L_{I}=L_{I}'$ and $M_{I}=-M_{I}'$. Neglecting spin-orbit coupling contribution implies invariance of the energy with respect to rotations of the atomic and magnetic system individually. Therefore $L_{I}=L_{I}'=L_{R}=0$ and $M_{I}=M_{I}'=M_{R}=0$ and the above relation reduces to,
\begin{equation}
\mathcal{C}_{l_{1}^{\prime}l_{2}^{\prime}L_{12}^{\prime}}^{m_{1}^{\prime}m_{2}^{\prime}M_{12}^{\prime}}\mathcal{C}_{L_{12}^{\prime}l_{3}^{\prime}0}^{M_{12}^{\prime}m_{3}^{\prime}0}
\mathcal{C}_{l_{1}l_{2}0}^{m_{1}m_{2}0}
\end{equation}
The length of vectors of atomic indices (i.e., $\bm{\mu}$, $\textbf{n}$, $\textbf{l}$ and $\textbf{m}$) is equal to the order $N$, while vectors of magnetic indices (i.e. $\textbf{n}'$, $\textbf{l}'$ and $\textbf{m}'$) have the length $N+1$, as can be seen in Fig.~\ref{fig:clbsch}.

\section{Optimized construction of basis and gradients}

The computational cost of the evaluation of the atomic basis can be reduced by exploiting 
\begin{equation}
Y_{l}^{-m}\left(\theta,\phi\right)=\left(-1\right)^{m}Y_{l}^{m \star}\left(\theta,\phi\right).
\end{equation}
for the computation of
\begin{align}
A_{i\mu_{i}n_{0}'l_{0}'-m_{0}'}^{(0)}&=(-1)^{m_{0}}A_{i\mu_{i}n_{0}'l_{0}'m_{0}'}^{(0)\star} \,, \\
A_{i\mu n00n'l'-m'} &= (-1)^{m'}A_{i\mu n00n'l'm'}^{\star}\,, \\
\label{eq:ayms1}
A_{i\mu nlmn'l'-m'} &= (-1)^{m+m'}A_{i\mu nl-mn'l'm'}^{\star} \,, \\
\label{eq:ayms2}
A_{i\mu nl-mn'l'-m'} &= (-1)^{m+m'}A_{i\mu nlmn'l'm'}^{\star}\,.
\end{align}
Similar optimizations were implemented for forces and magnetic gradients. In particular, the computational cost for the evaluation of the pseudo-force $\textbf{f}_{ik}$ in Eq.~(\ref{eq:pseudof}) can be reduced isolating the second body order contribution as follows
\begin{equation}\label{eq:force}
\begin{split}
&\textbf{f}_{ik} =\sum_{nn'l'm'}\omega_{i\mu_{k}n00n'l'm'}^{(1)}\nabla_{k}\phi_{\mu_{k}\mu_{i}n00n'l'm'}+\\
& \sum_{nlmn'l'm'}\omega_{i\mu_{k}nlmn'l'm'}^{(N>1)}\nabla_{k}\phi_{\mu_{k}\mu_{i}nlmn'l'm'}
\end{split}
\end{equation}
and considering that only the real part needs to be evaluated, therefore for the first term in the equation above one needs to calculate only the contribution for $m'\ge 0$
\begin{equation}\label{eq:decomp1}
\begin{split}
&\sum_{nn'l'm'}\omega_{i\mu_{k}n00n'l'm'}^{(1)}\nabla_{k}\phi_{\mu_{k}\mu_{i}n00n'l'm'}=\\
&\sum_{nn'l'}\Re\left(\omega_{i\mu_{k}n00n'l'0}^{(1)}\right)\Re\left( \nabla_{k}\phi_{\mu_{k}\mu_{i}n00n'l'0}  \right) + \\
&\sum_{nn'l'}\sum_{m'> 0}2\Re\left( \omega_{i\mu_{k}n00n'l'm'}^{(1)}\nabla_{k}\phi_{\mu_{k}\mu_{i}n00n'l'm'} \right).
\end{split}
\end{equation}
The second term of Eq.~(\ref{eq:force}) is split into five contributions by noting that the matrices of adjoints satisfy
\begin{equation}
\begin{split}
&\omega_{0-m'}=\left(-1\right)^{m'}\omega_{0m'}^{\star}\,,\\
&\omega_{-m0}=\left(-1\right)^{m}\omega_{m0}^{\star}\,,\\
&\omega_{-m-m'}=\left(-1\right)^{m+m'}\omega_{mm'}^{\star}\,,\\
&\omega_{m-m'}=\left(-1\right)^{m+m'}\omega_{-mm'}^{\star} \,.
\end{split}
\end{equation}
Therefore the second term of the force is decomposed as
\begin{equation}
\begin{split}
& \sum_{nlmn'l'm'}\omega_{i\mu_{k}nlmn'l'm'}\nabla_{k}\phi_{\mu_{k}\mu_{i}nlmn'l'm'} = \\
& \sum_{nln'l'}\Re\left(\omega_{i\mu_{k}nl0n'l'0}\right)\Re\left(\nabla_{k}\phi_{\mu_{k}\mu_{i}nl0n'l'0}\right) + \\
& \sum_{nln'l'}\sum_{m'> 0}2\Re\left( \omega_{i\mu_{k}nl0n'l'm'}\nabla_{k}\phi_{\mu_{k}\mu_{i}nl0n'l'm'} \right) + \\ & \sum_{nln'l'}\sum_{m> 0}2\Re\left( \omega_{i\mu_{k}nlmn'l'0}\nabla_{k}\phi_{\mu_{k}\mu_{i}nlmn'l'0} \right) + \\
&\sum_{nln'l'}\sum_{m>0,m'> 0}2\Re\left( \omega_{i\mu_{k}nlmn'l'm'}\nabla_{k}\phi_{\mu_{k}\mu_{i}nlmn'l'm'} \right) +\\ & \sum_{nln'l'}\sum_{m< 0,m'>0}2\Re\left( \omega_{i\mu_{k}nlmn'l'm'}\nabla_{k}\phi_{\mu_{k}\mu_{i}nlmn'l'm'} \right).
\end{split}
\end{equation}
In this way one saves 25$\%$ of computational time. 
Similar considerations can be drawn for the magnetic gradients. In particular, the pseudo-magnetic gradient $\textbf{t}_{ki}$ of Eq.~(\ref{eq:tki})
can be decomposed as the pseudo-atomic force of Eq.~(\ref{eq:force}), where the expression of the adjoints is equivalent, while for the pseudo-magnetic gradient $\textbf{t}_{k}$ we need to consider the additional adjoint $\omega_{k\mu_{k}n'l'm'}^{(0)}$ of Eq.~(\ref{eq:tk0}) that can be reduced to the real part using Eq.~(\ref{eq:decomp1}). 

\section{Workflow of the magnetic ACE implementation}

The following workflow was used for the evaluation of magnetic ACE atomic energy and atomic and magnetic gradients,
\begin{enumerate}
	\item{Generation of possible combinations of atomic and magnetic basis functions allowed by symmetry requirements (rotational and inversion invariance), lexicographical order and condition of linear independence (enforced with Singular Value Decomposition);}
	\item{Calculation of the onsite atomic basis $A_{i\mu_{i}n_{0}'l_{0}'m_{0}'}^{(0)}$ using radial functions and spherical harmonics of the atom $i$;}
	\item{Loop over the neighbors $j$ of atom $i$ to calculate the atomic basis $A_{i\mu nlmn'l'm'}$;}
	\item{Loop over basis function indices to calculate products of the atomic basis for each body order (i.e. Eq.~(\ref{eq:fgeneral})) and the quantities $dA^{(s)}_{i\bm{\mu}\textbf{n}\textbf{l}\textbf{m}\textbf{n}'\textbf{l}'\textbf{m}'}$ and $dA^{(0)}_{i\bm{\mu}\textbf{n}\textbf{l}\textbf{m}\textbf{n}'\textbf{l}'\textbf{m}'}$ of Eqs.~(\ref{eq:dbs}) and (\ref{eq:db0});}
	\item{At this point the single atom energies can be evaluated;}
	\item{The matrix $\Theta^{(N)}_{\mu_{i}\bm{\mu}\textbf{n}\textbf{l}\textbf{n}'\textbf{l}'}$ is determined from the expansion coefficients and then the adjoints of Eqs.~(\ref{eq:adj}) and (\ref{eq:adj0}) are calculated;}
	\item{The atomic and magnetic forces can be finally assembled.}
\end{enumerate}
The final computational cost for the determination of the single atom energies $\varepsilon_{i}$ (and in general of any scalar) depends on the evaluation of the onsite $A^{(0)}$ basis, the atomic basis $A$ and on the computation of the $B$ basis. The scaling is $O\left(\#A^{(0)}\right)$, $O\left(N_{c}\cdot\#A\right)$ and $O\left(\left(\left(N_{max}+1)+\#\varepsilon)\cdot\#B\right)\right)\right)$ respectively, where $N_{c}$ is the number of neighbors within the chosen cutoff radius and $N_{max}$ is the maximum order.  

\section{Tuning of the integration radius}

\begin{figure}[h]
	\centering
	\includegraphics[width=12.0cm]{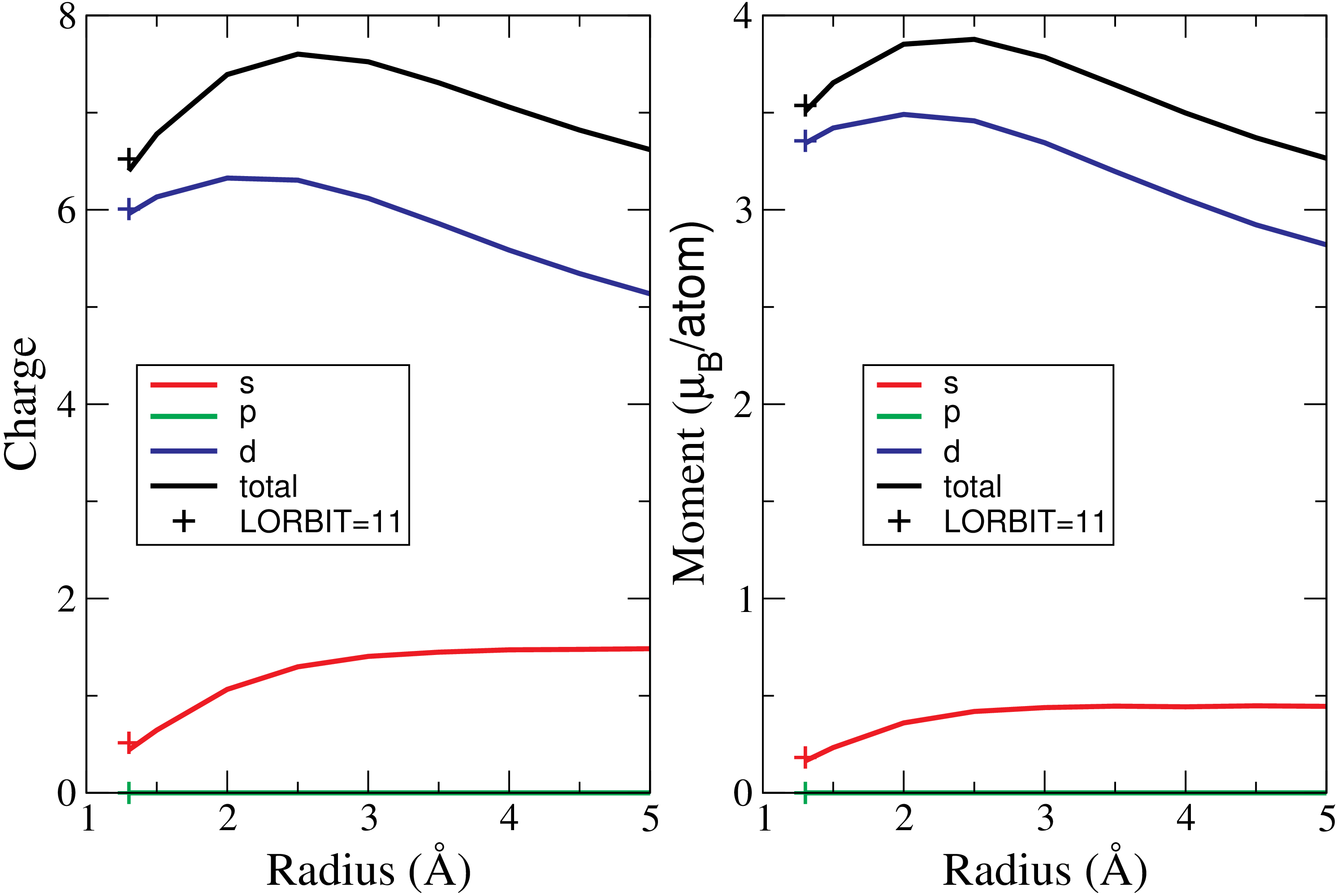}
	\caption{Charge and magnetic moment per atom vs integration radius for a single iron atom in a large supercell. The atom has 8 valence electrons and a magnetic moment equal to $4 \mu_{B}$. The crosses mark the default integration  radius RWIGS of 1.302 $\angstrom$ of the pseudopotential.}
	\label{fig:free_ws}
\end{figure}

The parameter RWIGS in VASP can be adjusted for bulk phases to the corresponding Wigner-Seitz radius. 
However, this procedure is not easily applicable to distorted structures. 
Nevertheless, the atomic magnetic moment (and also the charge associated with an atom) depends on the value of the integration radius. 
In order to illustrate this point, we show in Fig.~\ref{fig:free_ws} the charge and the magnetic moment vs the integration radius for a single Fe atom in a large supercell. The results are obtained by projecting s, p and d atomic-like orbitals on the self-consistent charge density. 
Both the charge and the magnetic moment exhibit  a considerable variation with respect to the integration radius, where the d electrons account for the largest contribution. For this orbital projection, the most appropriate integration radius for a free atom corresponds to about 2.5\,$\angstrom$, which results in a charge of almost 8 valence electrons and an atomic magnetic moment of almost 4\,$\mu_B$.

\begin{figure}[h]
	\centering
	\includegraphics[width=10.0cm]{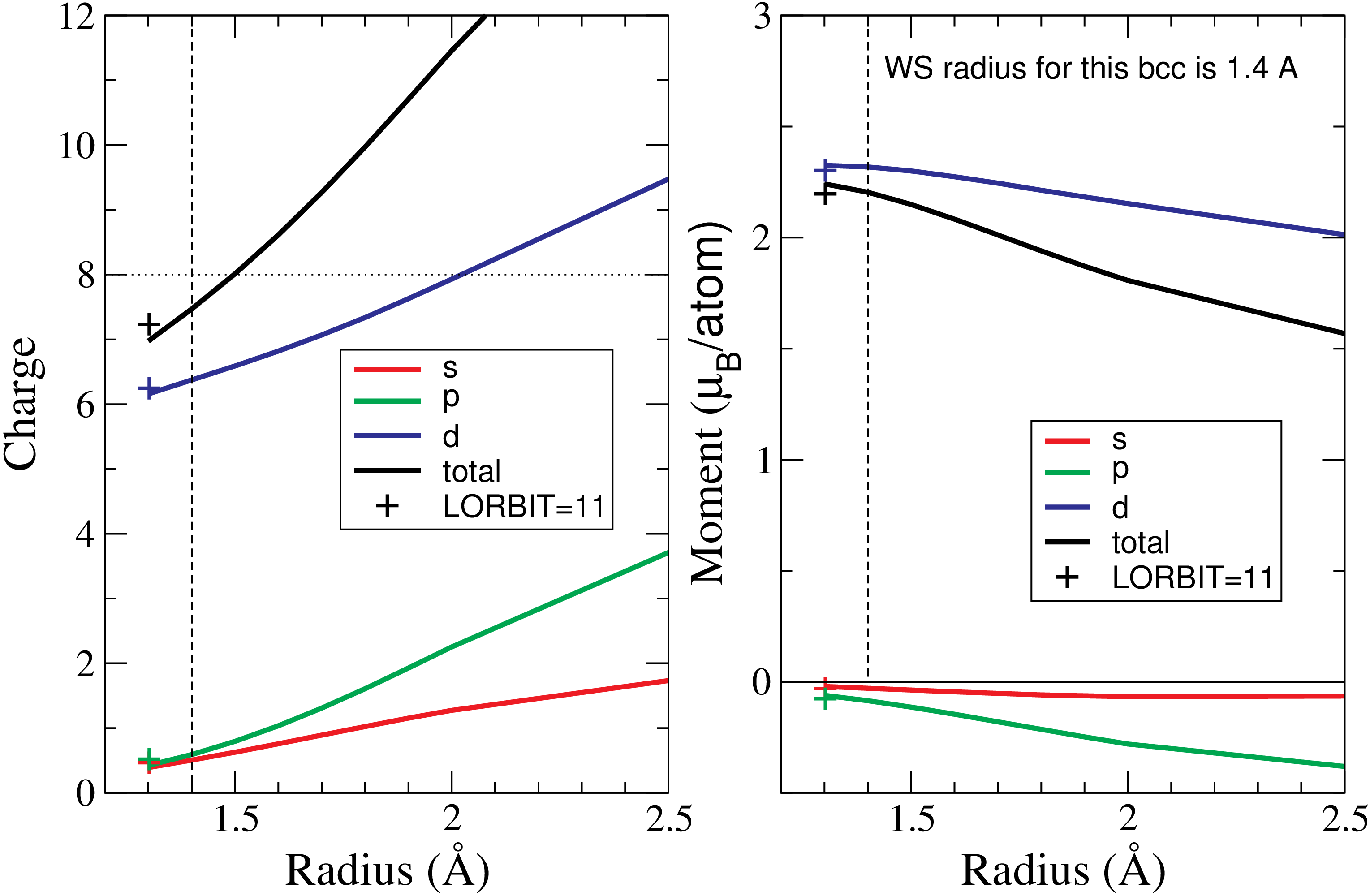}\\
	\includegraphics[width=10.0cm]{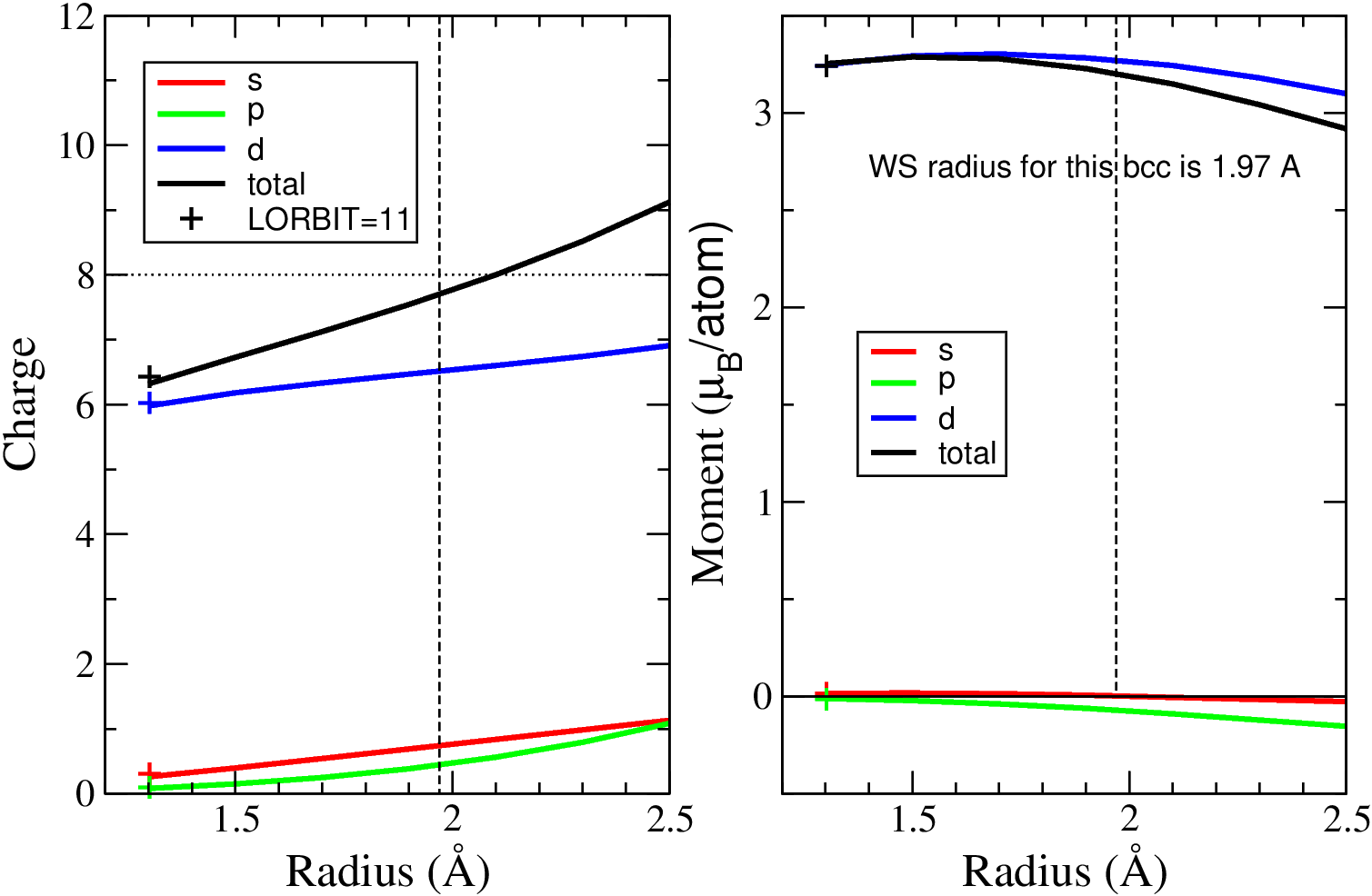}
	\caption{Charge and magnetic moment per atom vs integration radius for the FM bcc Fe with lattice parameter equal to 2.85 $\angstrom$ (top panels) and 4.0 $\angstrom$ (bottom panels).  The crosses mark the default integration radius RWIGS radius of 1.302 $\angstrom$ from the pseudopotential. The vertical dashed lines mark the Wigner-Seitz radius for the corresponding volume.}
	\label{fig:bcc_ws_2_85}
\end{figure}

Figure~\ref{fig:bcc_ws_2_85} shows similar dependencies for the FM bcc structure with lattice parameters equal to 2.85 and 4.0\,$\angstrom$, respectively. It is clear from these two plots that in the case of the bulk bcc phase, and generally for other bulk structures, the magnitude of the magnetic moment does not change much for moderate changes of the Wigner-Seitz radius, namely, between the default pseudopotential value of 1.302\,$\angstrom$ and the actual Wigner-Seitz radius corresponding to the given volume. Therefore, to avoid any ambiguity, we decided to use the same (default) integration radius in all calculations to obtain consistent values of magnetic moments for arbitrary configuration.

\section{Issues with the convergence of the penalty energy term for constrained DFT calculations}

\begin{figure}
    \includegraphics[width=12.0cm]{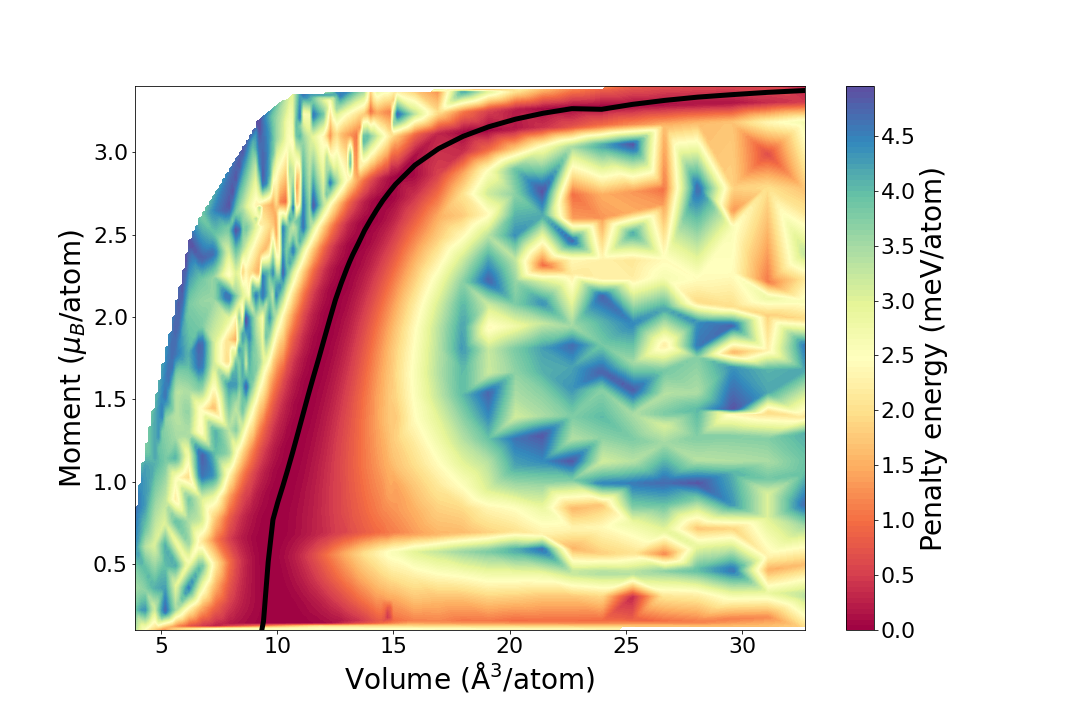}
    \caption{Map of the penalty energy for bcc AFM. The equilibrium configurations are marked by the black curve.}
    \label{fig:penalty}
\end{figure}

A representative example showing a map of the penalty energy for antiferromagnetic (AFM) bcc as function of volume and constrained magnetic moment is shown in Fig.~\ref{fig:penalty}. One can see that the region of lowest penalty energy ($\sim$1\,meV/atom) is within a curved trench (dark red color) close to the lowest energy magnetic configurations as a function of volume (black curve). Away from this region the penalty energy increases up to to about 4.5\,meV/atom for large-volume/small-moment configurations (the lower right plateau region) or more for configurations characterized by large magnetic moments and small volumes (the upper left blue region). Therefore, for our training, we restricted the dataset to regions characterized by low magnitudes of the penalty energy, i.e. by excluding both large and low volume data.

\section{Supplementary plots}

\begin{figure}
     \includegraphics[width=12.0cm]{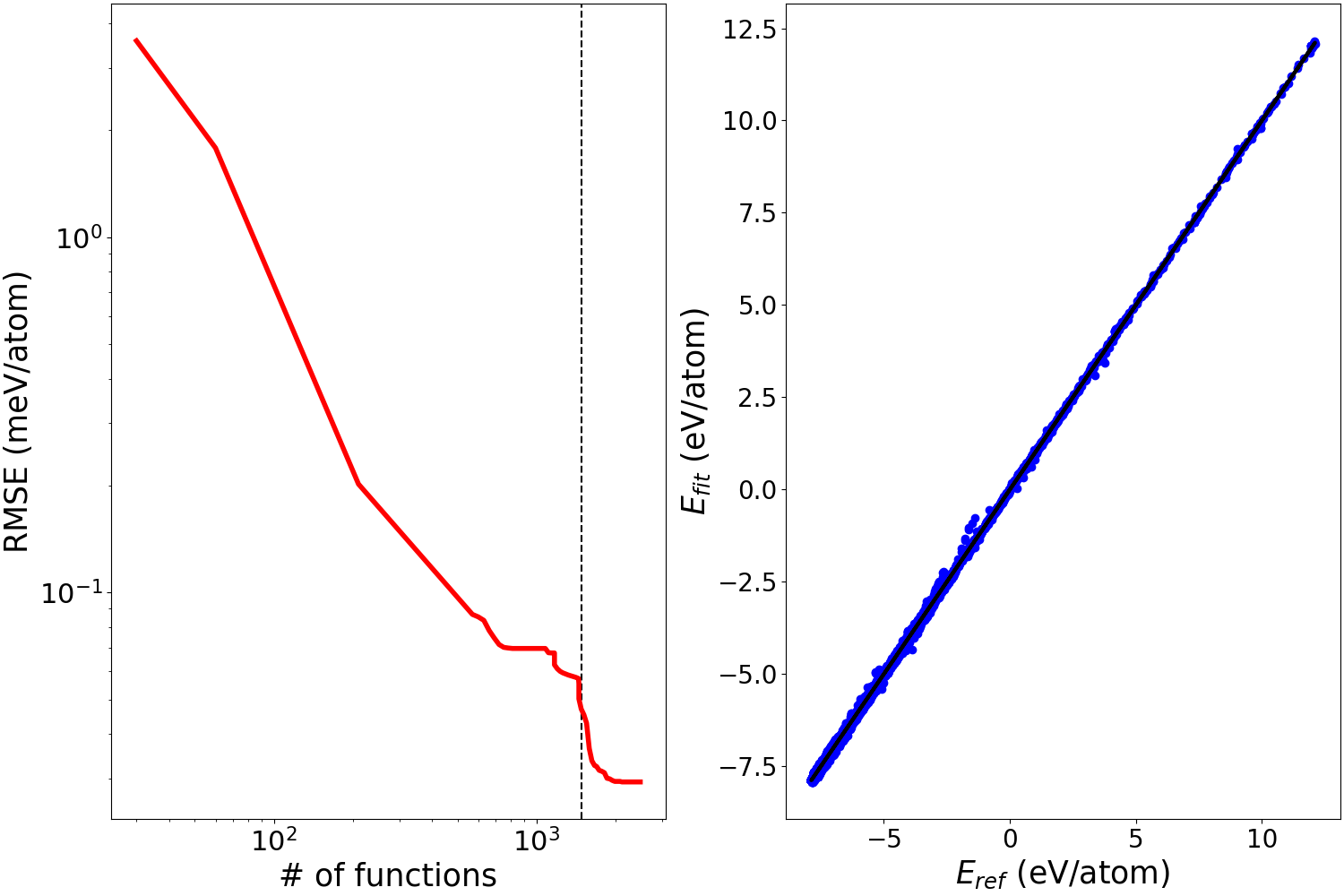}
    \caption{Left panel: learning curve for the fit of a prototypical database with 4592 structures. At the vertical three-body angular magnetic terms were added to the two-body magnetic interactions. Right panel: DFT reference ($E_{ref}$) vs ACE ($E_{fit}$) energies.}
    \label{fig:rmse}
\end{figure}

The learning curve for the magnetic ACE training is shown in Fig.~\ref{fig:rmse} together with the correlation plot of predicted vs reference energies.

\begin{figure}[h]
	\centering
	\includegraphics[width=12.0cm]{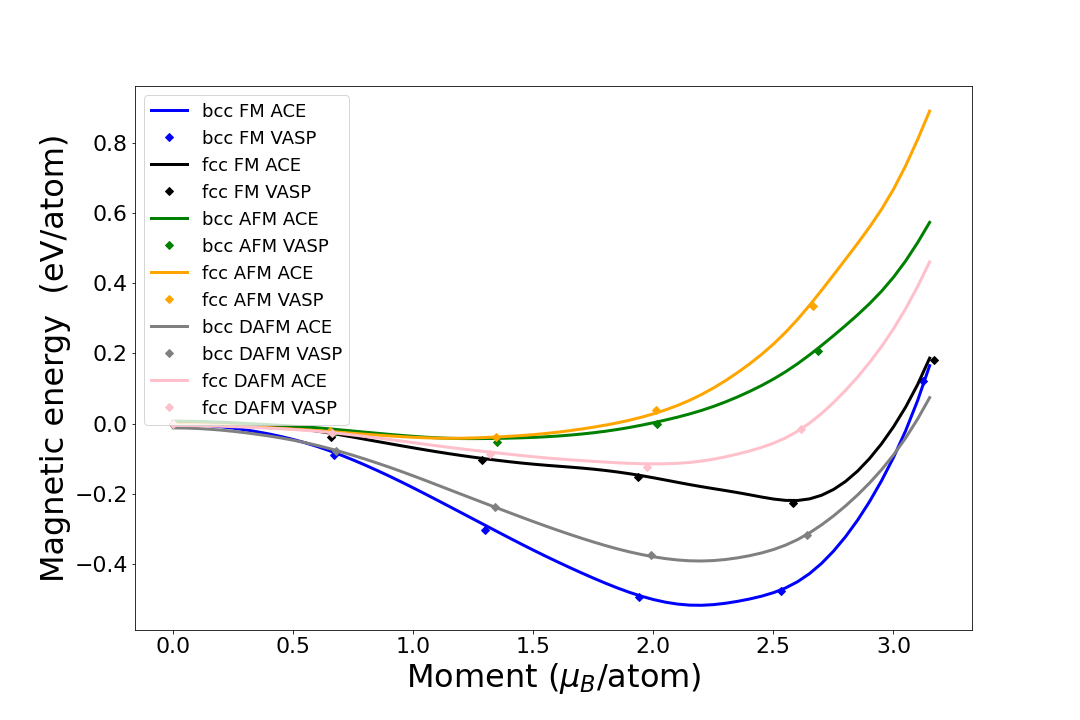}
	\caption{Magnetic energy vs magnetic moment magnitude curves for the different magnetic phases of bcc and fcc together with the DFT reference.}
	\label{fig:em_mace1}
\end{figure}

\begin{figure}[h]
	\centering
	\includegraphics[width=12.0cm]{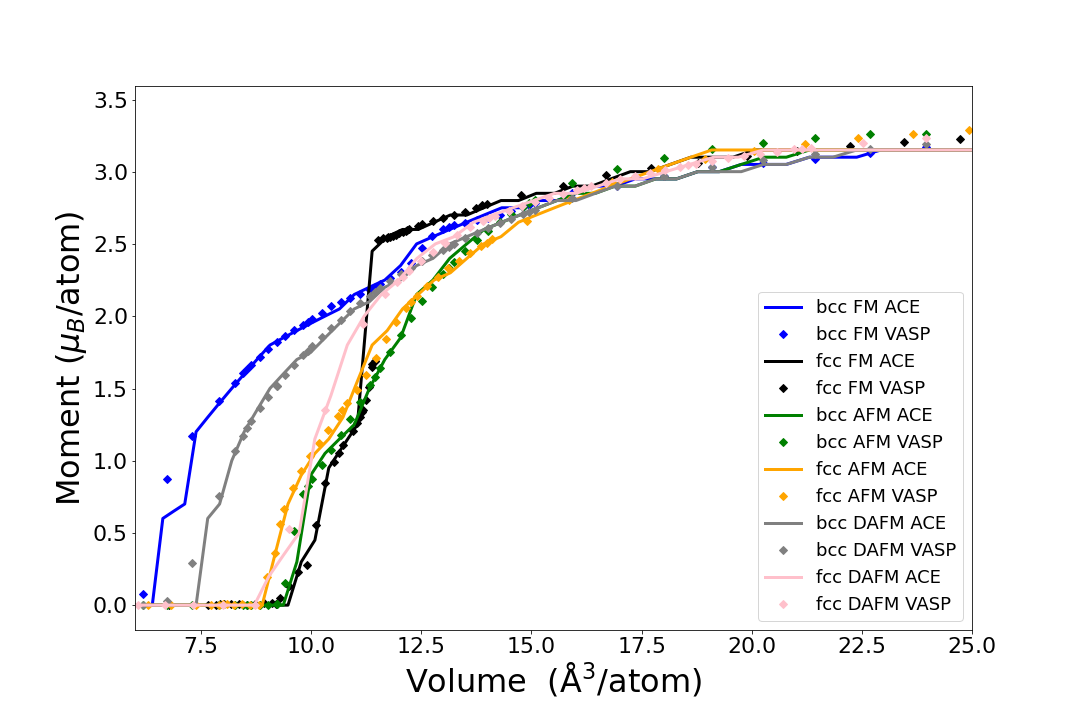}
	\caption{Magnetic moment magnitude vs volume curves for the magnetic phases of bcc and fcc with the corresponding reference DFT data.}
	\label{fig:vm_mace1}
\end{figure}

\begin{figure}[h]
	\centering
	\includegraphics[width=14.0cm]{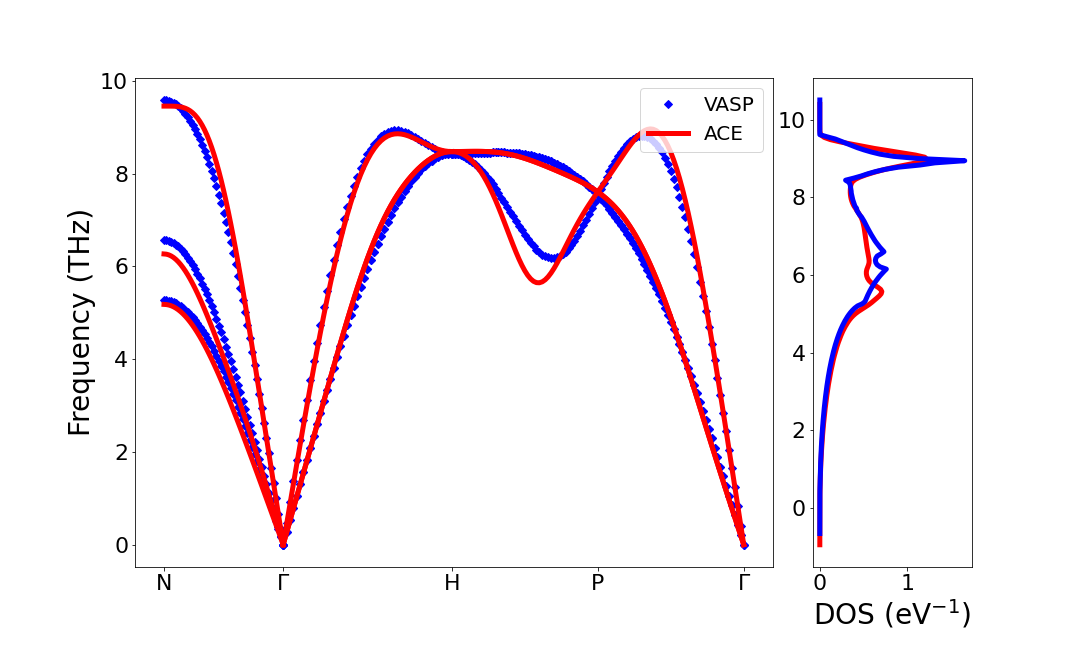}
	\caption{Phonon spectra for bcc FM at the equilibrium volume and magnetic moment.}
	\label{fig:phonbcc_mace}
\end{figure}

The equilibrium curves for the magnetic energy vs magnetic moment magnitude are displayed in Fig.~\ref{fig:em_mace1}, where the different magnetic phases of bcc and fcc show varying degrees of localization. For instance, in the case of bcc FM the moment at the equilibrium volume is sufficiently localized if compared to fcc FM. The latter shows indeed a more pronounced itinerant character, leading to energetically favorable LSFs at finite temperatures and larger spread in the distribution of the moments~\cite{alling2016strong,dong2017longitudinal}. The related equilibrium moment magnitude vs volume curves are given in Fig.~\ref{fig:vm_mace1}, where the equilibrium state for all the magnetic phases changes rapidly from the NM to the magnetic solution in a range of volumes from $\approx 6 \angstrom^{3}$/atom to 10 $\angstrom^{3}$/atom and stabilizes around 12 $\angstrom^{3}$/atom increasing then more gently up to the free atom limit. In the case of fcc FM the high- and low-spin states are also visible at the corresponding volumes. The non-smooth behavior of these curves is determined by the details of the integrated density of states (DOS) for the spin up and down channels, which are given by the competition between electronic kinetic and Coulomb energy. \\
The FM phonon spectrum at the equilibrium magnetic moment is displayed, together with the corresponding DOS, in Fig.~\ref{fig:phonbcc_mace}, where also an excellent agreement is obtained with the reference DFT spectrum. \\

\section{Specific model fitted on defects only}

\begin{figure}[h]
	\centering
	\includegraphics[width=12.0cm]{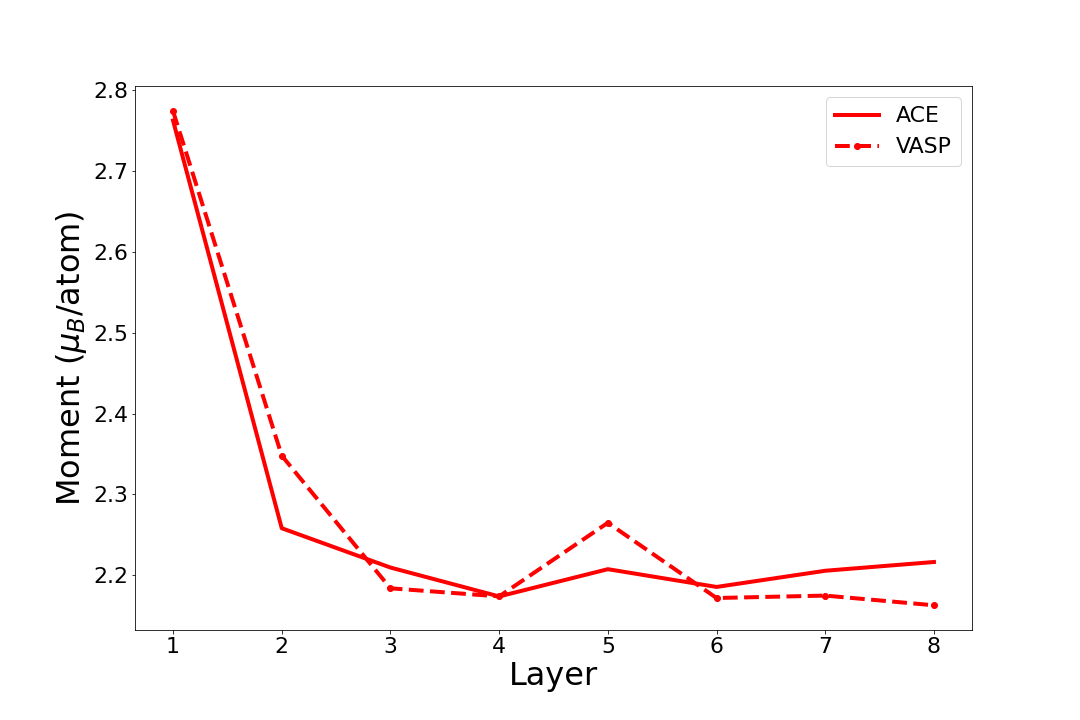}
	\caption{Magnetic moment as a function of the layer depth for the (112) surface with relaxed both atomic positions and magnetic moments.}
	\label{fig:surfaces}
\end{figure}

In this section we discuss the predictions obtained by fitting ACE on defects and bcc FM only. The chosen cutoff is equal to 6.0 $\angstrom$ and RMSE are 14 meV for energies and 153 meV/$\angstrom$ for forces.
A list of defect formation energies is shown in Tab.~\ref{table:vac} together with the reference DFT data. The agreement is good for all the defects properties despite the lack of sampling of the magnetic DOF due to convergence problems of the constrained local moment method in this specific case. In Tab.~\ref{table:vacmom} the predicted distribution of the magnetic moments around the monovacancy is shown with the reference DFT data. In the particular case of the (112) surface we also compared the values of the relaxed moments as a function of the layer depth, as shown in Fig.~\ref{fig:surfaces}, to the reference. The agreement for this particular surface is within an acceptable range of $\pm$ 0.1 $\mu_{B}$/atom. The predictions for the 1/2[111] screw dislocation core structure and $\gamma$-surface are identical to the results obtained with the potential presented in the main text. 

\begin{table*}[t]
	\centering
	\setlength{\tabcolsep}{10pt} 
    \renewcommand{\arraystretch}{1.5}
    \begin{tabular}{|l|l | r |}
        \hline
            & ACE & DFT \\ [0.5ex] 
        \hline
         $E_{vac}$ (eV) & 2.21 & 2.17 \\ 
          $E_{m}$ & 0.65 & 0.67\\
        $E^{100}$ & 5.44 & 5.31  \\
        $E^{110}$ & 4.84 & 4.88  \\
        $E^{111d}$ & 4.82 & 4.85  \\
        $E^{111c}$ & 4.82 & 4.85  \\
        $E^{100}$ (mJ/$m^{2}$) & 2587 & 2512 \\
        $E^{110}$  & 2342 & 2431 \\
        $E^{111}$ & 2857 & 2680 \\
        $E^{112}$ & 2700 & 2749 \\
        \hline
        \end{tabular}
        \caption{List of defect formation energies as predicted by ACE when both atoms and magnetic moments are relaxed in comparison with our reference DFT values. In the case of interstitials and surfaces the predicted formation energies are calculated by relaxing only the positions and keeping the moments fixed to the equilibrium bulk value.}
    \label{table:vac}
\end{table*}

\begin{table*}[t]
	\centering
    \begin{tabular}{|c | l  | r |}
        \hline
         Site & ACE & DFT \\ [0.5ex] 
        \hline
         & & \\ 
         1NN & 2.34 & 2.42\\
         2NN & 2.27 & 2.08\\
         3NN & 2.28 & 2.18\\
         & & \\ 
        \hline
        \end{tabular}
        \caption{Values of the magnetic moments in $\mu_{B}$ for the $1^{st}$, $2^{nd}$ and $3^{rd}$ nearest-neighbor of the vacancy site. }
    \label{table:vacmom}
\end{table*}

\end{document}